\documentclass[prc,twocolumn,showpacs]{revtex4}

\usepackage{graphicx}
\usepackage{dcolumn}
\usepackage{bm}

\begin{document}

\title{Beta-decay properties of neutron-rich Ge, Se, Kr, Sr, Ru, and Pd isotopes
from deformed quasiparticle random-phase approximation}

\author{P. Sarriguren}
\email{p.sarriguren@csic.es}
\affiliation{
Instituto de Estructura de la Materia, IEM-CSIC, Serrano
123, E-28006 Madrid, Spain}

\date{\today}

\begin{abstract}

$\beta$-decay properties of even and odd-$A$ neutron-rich Ge, Se, Kr, Sr, Ru, 
and Pd isotopes involved in the astrophysical rapid neutron capture process 
are studied within a deformed proton-neutron quasiparticle random-phase 
approximation. 
The underlying mean field is described self-consistently from deformed 
Skyrme Hartree-Fock calculations with pairing correlations.
Residual interactions in the particle-hole and particle-particle channels
are also included in the formalism. 
The isotopic evolution of the various nuclear equilibrium shapes and the
corresponding charge radii are investigated in all the isotopic chains.
The energy distributions of the Gamow-Teller strength as well as the 
$\beta$-decay half-lives are discussed and compared with the available
experimental information. It is shown that nuclear deformation plays a 
significant role in the description of the decay properties in this mass
region. Reliable predictions of the strength distributions are essential 
to evaluate decay rates in astrophysical scenarios.

\end{abstract}

\pacs{21.60.Jz, 23.40.Hc,  27.60.+j, 26.30.-k}

\maketitle

\section{Introduction}

The rapid structural changes occurring in the ground state and low-lying 
collective excited states of neutron-rich nuclei in the mass region 
$A= 80-128$ have been extensively studied both theoretical and 
experimentally (see e.g. \cite{wood92,heyde11} and references therein). 
From the theoretical side, the equilibrium nuclear shapes in this mass 
region have been shown to suffer rapid changes as a function 
of the number of nucleons with competing spherical, axially symmetric 
prolate and oblate, and triaxial shapes at close energies. Both 
relativistic \cite{xiang12,mei12} and nonrelativistic 
\cite{bonche85,bender08,bender09,rayner1,rayner2} approaches agree 
in the general description of the nuclear structural evolution in 
this mass region, which is supported experimentally by spectroscopic 
studies \cite{wu04,urban04,sumikama11}, $2^+$ lifetime measurements
\cite{mach,goodin,urban} and quadrupole moments for rotational 
bands \cite{urban}, as well as by laser spectroscopy measurements 
\cite{cheal10}.

However, the nuclear structure richness is not the only attractive 
feature characterizing these nuclei. Another remarkable property
of nuclei in this mass region is that they are involved in the 
astrophysical rapid neutron capture process ({\it r} process), which 
is considered to be one of the main nucleosynthesis mechanisms 
leading to the production of heavy neutron-rich nuclei in the 
universe \cite{bbhf,cowan91}.
The {\it r}-process nucleosynthesis involves many neutron-rich unstable 
isotopes, whose neutron capture rates, masses, and $\beta$-decay 
half-lives ($T_{1/2}$) are crucial quantities to understand the possible 
{\it r}-process paths, the isotopic abundances, and the time scales of 
the process \cite{cowan91,kra93,nishi12}. Although much progress has 
been done measuring masses (see for example the Jyv\"askyl\"a mass 
database \cite{jyvaskyla}) and half-lives \cite{pereira,nishimura11,quinn12}, 
most of the nuclear properties of relevance for the {\it r} process are 
experimentally unknown due to their extremely low production yields in 
the laboratory. Therefore, reliable nuclear physics models are required 
to simulate properly the {\it r} process.

The quasiparticle random-phase approximation (QRPA) is considered a 
well suited model to describe medium-mass open-shell nuclear
properties and specifically $\beta$-decay properties. QRPA calculations 
for neutron-rich nuclei have been carried out within different spherical 
formalisms, such as Hartree-Fock-Bogoliubov (HFB) \cite{engel}, continuum 
QRPA with density functionals \cite{borzov3}, and relativistic mean field 
approaches \cite{niksic2005}.
However, the mass region we are dealing with requires nuclear deformation 
as a relevant degree of freedom to characterize the nuclear structure 
involved in the calculation of the $\beta$-strength functions. 
The deformed QRPA formalism was developed in Refs. 
\cite{moller1,moller3,homma,hir2}, using
phenomenological mean fields. A Tamm-Dancoff approximation with Sk3 
interaction was also implemented in Ref. \cite{hamamoto}.
More recently, deformed QRPA calculations using deformed Woods-Saxon 
potentials and realistic CD-Bonn residual forces have been performed 
in \cite{fang13,ni14}. 
First-forbidden transitions were also considered in those references,
showing that their effect in this mass region can be neglected. Various 
self-consistent deformed QRPA calculations to describe the $\beta$-decay 
properties, either with Skyrme \cite{yoshida13} or Gogny \cite{peru14} 
interactions are also available in the literature.

In Refs. \cite{sarripere,sarri14} the decay properties of neutron-rich Zr 
and Mo isotopes were studied within a deformed proton-neutron QRPA 
based on a self-consistent Hartree-Fock (HF) mean field formalism with 
Skyrme interactions and pairing correlations in BCS approximation. 
Residual spin-isospin interactions were also included in the 
particle-hole and particle-particle channels \cite{sarri1,sarri2}.
In this work this study is extended to the neighboring regions 
including even and odd-$A$ neutron-rich Ge, Se, Kr, Sr, Ru, and Pd 
isotopes. These calculations are timely because they address a mass 
region which is at the borderline of present experimental capabilities 
for measuring half-lives at MSU and 
RIKEN \cite{pereira,nishimura11,quinn12}. 
In addition, theoretical calculations can be tested with the available 
experimental information on half-lives providing simultaneously
predictions for the underlying Gamow-Teller strength distributions 
and for the half-lives of more exotic nuclei not yet measured.
Finally, this more comprehensive study allows one to judge better
the extent to which the method is able to describe the decay properties
of nuclei in a wider mass region that includes spherical, well deformed,
and weakly deformed transitional isotopes, as well as isotopes
exhibiting shape coexistence. Therefore, the theoretical method will 
be tested over a rich set of different nuclear structures that will 
reveal the limitations of the model.

The paper is organized as follows. In Sec. \ref{sec2} a review of the 
theoretical formalism used is introduced. Section \ref{results} 
contains the results obtained for the potential energy curves (PEC), 
Gamow-Teller (GT) strength distributions, and $\beta$-decay 
half-lives, which are compared with the experimental data. 
Section IV summarizes the main conclusions.

\section{Theoretical Formalism}
\label{sec2}

A summary of the theoretical framework used in this paper to describe 
the $\beta$-decay properties in neutron-rich isotopes is shown in this
section. More details of the formalism can be found 
elsewhere \cite{sarri1,sarri2}.
The method starts from a self-consistent calculation of the mean field
by means of a deformed Skyrme Hartree-Fock procedure with pairing 
correlations in BCS approximation. 
Single-particle energies, wave functions, and occupation amplitudes are 
generated from this mean field. The Skyrme interaction SLy4 \cite{sly4} 
is used as a representative of modern Skyrme forces. It has 
been very successful describing nuclear properties all along the nuclear 
chart and has been extensively studied \cite{bender08,bender09,stoitsov}.

The solution of the HF equation, assuming time reversal and axial 
symmetry, is found by using the formalism developed in 
Ref. \cite{vautherin}. The single-particle wave functions are expanded 
in terms of the eigenstates of an axially symmetric harmonic oscillator 
in cylindrical coordinates, using twelve major shells. 
The pairing gap parameters for protons and neutrons in the BCS 
approximation are determined phenomenologically from the odd-even 
mass differences \cite{audi12}. 
In a further step, constrained HF calculations with a quadratic 
constraint are performed to construct the PECs, analyzing the 
nuclear binding energies in terms of the  quadrupole deformation 
parameter $\beta$. 
Calculations for GT strengths are performed subsequently for the
various minima in the energy curves indicating the equilibrium 
shapes of each nucleus.
Since decays connecting different shapes are disfavored, similar 
shapes are assumed for the ground state of the parent nucleus and 
for all populated states in the daughter nucleus. The validity of this 
assumption was discussed for example in Refs. \cite{moller1,homma}. 

To describe GT transitions, a separable spin-isospin residual 
interaction in the particle-hole (ph) and particle-particle (pp) 
channels is added to the mean field and treated in a deformed 
proton-neutron QRPA
\cite{moller1,moller3,homma,hir2,hamamoto,sarri1,sarri2,nabi}.
An optimum set of coupling strengths could be chosen following a 
case by case fitting procedure and one will finally get different 
answers depending on the nucleus, shape, and Skyrme force. However, 
since the purpose here is to test the ability of QRPA to account 
for the GT strength distributions in this mass region with as few 
free parameters as possible, the same coupling strengths are used
for all the nuclei considered in this paper, which are 
taken from previous works \cite{sarripere,sarri14}. We use 
$\chi ^{ph}_{GT}=0.15$ MeV and $\kappa ^{pp}_{GT} = 0.03$ MeV for 
the residual interaction in the ph and pp channels, respectively.

The sensitivity of the GT strength distributions to the various 
ingredients contributing to the deformed QRPA calculations, namely to 
the nucleon-nucleon effective force, to deformation, to pairing 
correlations, and to residual interactions, have been investigated 
in the past \cite{sarri1,sarri2,sarri3,sarri4,sarriwp}. In this
work the most reasonable choices found in those references are used.
Summarizing the various sensitivities, the conclusion is that the main 
features of the GT strength distributions are in general very 
robust against the Skyrme force used, showing some more sensitivity 
in the spherical cases, where the location of the
single-particle energies is more critical to determine
the excitation energies of the GT transitions. Deformation has
been shown to be an important issue to describe the profiles
of the GT strength distributions. First, because the degeneracy
of the spherical shells is broken making the GT strength 
distributions more fragmented than the corresponding spherical 
ones. Secondly, because the energy levels of deformed orbitals
cross each other in a way that depends on the magnitude of the
quadrupole deformation as well as on the oblate or prolate 
character. This level crossing may lead in some instances to
sizable differences in the GT profiles, a fact that has been
exploited to learn about the nuclear shape from the measured
$\beta$-decay
pattern \cite{exppoirier,expnacher,expperez}. Pairing correlations 
are also important to describe nuclei out of closed shells. Their
influence on the GT profiles was studied in Ref. \cite{sarri2}, 
concluding that the main effect is to decrease slightly the strength 
at low energies and to create new transitions, mainly at high 
energies, that are forbidden in the absence of such correlations.
The effect of the ph and pp residual interactions is also
well known. The repulsive ph interaction redistributes the GT 
strength by shifting it to higher excitation energies causing a 
displacement of the GT resonance. It also reduces somewhat the 
total strength. The attractive pp interaction moves the strength 
to lower energies. Its effect on the GT resonance is in general 
negligible, but nevertheless, the changes induced in the
low-energy region are of great relevance in the calculation of the 
$\beta$-decay half-lives, which are only sensitive to the strength 
contained in the energy region below the $Q$-energy window.

The GT transition amplitudes in the intrinsic frame connecting the 
ground state $| 0^+\rangle $ of an even-even nucleus to one phonon 
states in the daughter nucleus $|\omega_K \rangle \, (K=0,1) $ are 
found to be

\begin{equation}
\left\langle \omega _K | \sigma _K t^{\pm} | 0 \right\rangle =
\mp M^{\omega _K}_\pm \, ,
\label{intrinsic}
\end{equation}
where
\begin{eqnarray}
M_{-}^{\omega _{K}}&=&\sum_{\pi\nu}\left( q_{\pi\nu}X_{\pi
\nu}^{\omega _{K}}+ \tilde{q}_{\pi\nu}Y_{\pi\nu}^{\omega _{K}}
\right) , \\
M_{+}^{\omega _{K}}&=&\sum_{\pi\nu}\left(
\tilde{q}_{\pi\nu} X_{\pi\nu}^{\omega _{K}}+
q_{\pi\nu}Y_{\pi\nu}^{\omega _{K}}\right) \, ,
\end{eqnarray}
with
\begin{equation}
\tilde{q}_{\pi\nu}=u_{\nu}v_{\pi}\Sigma _{K}^{\nu\pi },\ \ \
q_{\pi\nu}=v_{\nu}u_{\pi}\Sigma _{K}^{\nu\pi},
\label{qs}
\end{equation}
in terms of the occupation amplitudes for neutrons and protons $v_{\nu,\pi}$   
($u^2_{\nu,\pi}=1-v^2_{\nu,\pi}$) and the matrix elements of the spin operator, 
$\Sigma _{K}^{\nu\pi}=\left\langle \nu\left| \sigma _{K}\right| 
\pi\right\rangle $, connecting proton and neutron single-particle states, 
as they come out from the HF+BCS calculation. $X_{\pi\nu}^{\omega _{K}}$ and 
$Y_{\pi\nu}^{\omega _{K}}$ are the forward and backward amplitudes of the 
QRPA phonon operator, respectively. 

Once the intrinsic amplitudes in Eq. (\ref{intrinsic}) are calculated, 
the GT strength $B_{\omega}(GT^\pm)$ in the laboratory system for a 
transition  $I_iK_i (0^+0) \rightarrow I_fK_f (1^+K)$ can be obtained as
\begin{eqnarray}
B_{\omega}(GT^\pm )& =& \sum_{\omega_{K}} \left[ \left\langle \omega_{K=0}
\left| \sigma_0t^\pm \right| 0 \right\rangle ^2 \delta (\omega_{K=0}-
\omega ) \right.  \nonumber  \\
&& \left. + 2 \left\langle \omega_{K=1} \left| \sigma_1t^\pm \right|
0 \right\rangle ^2 \delta (\omega_{K=1}-\omega ) \right] \, ,
\label{bgt}
\end{eqnarray}
in $[g_A^2/4\pi]$ units. To obtain this expression, the initial and
final states in the laboratory frame have been expressed in terms of
the intrinsic states using the Bohr-Mottelson factorization \cite{bm}.

The specific treatment of odd-$A$ systems has been described 
\cite{hir2,sarri4} by considering two types of GT contributions. One 
type is due to phonon excitations in which the odd nucleon acts only 
as a spectator. The transition amplitudes in the intrinsic frame 
are in this case basically the same as in the even-even case, but 
with the blocked spectator excluded from the calculation. The other 
type of transitions involves the odd nucleon and is treated 
perturbatively by taking into account phonon correlations to first 
order in the quasiparticle transitions.
The excitation energies of the GT states with respect to the ground 
state in the daughter nuclei have been discussed in Ref. \cite{sarri4}
for both types of transitions in terms of the QRPA phonon energy and
the quasiparticle energies.

\begin{figure}[ht]
\centering
\includegraphics[width=85mm]{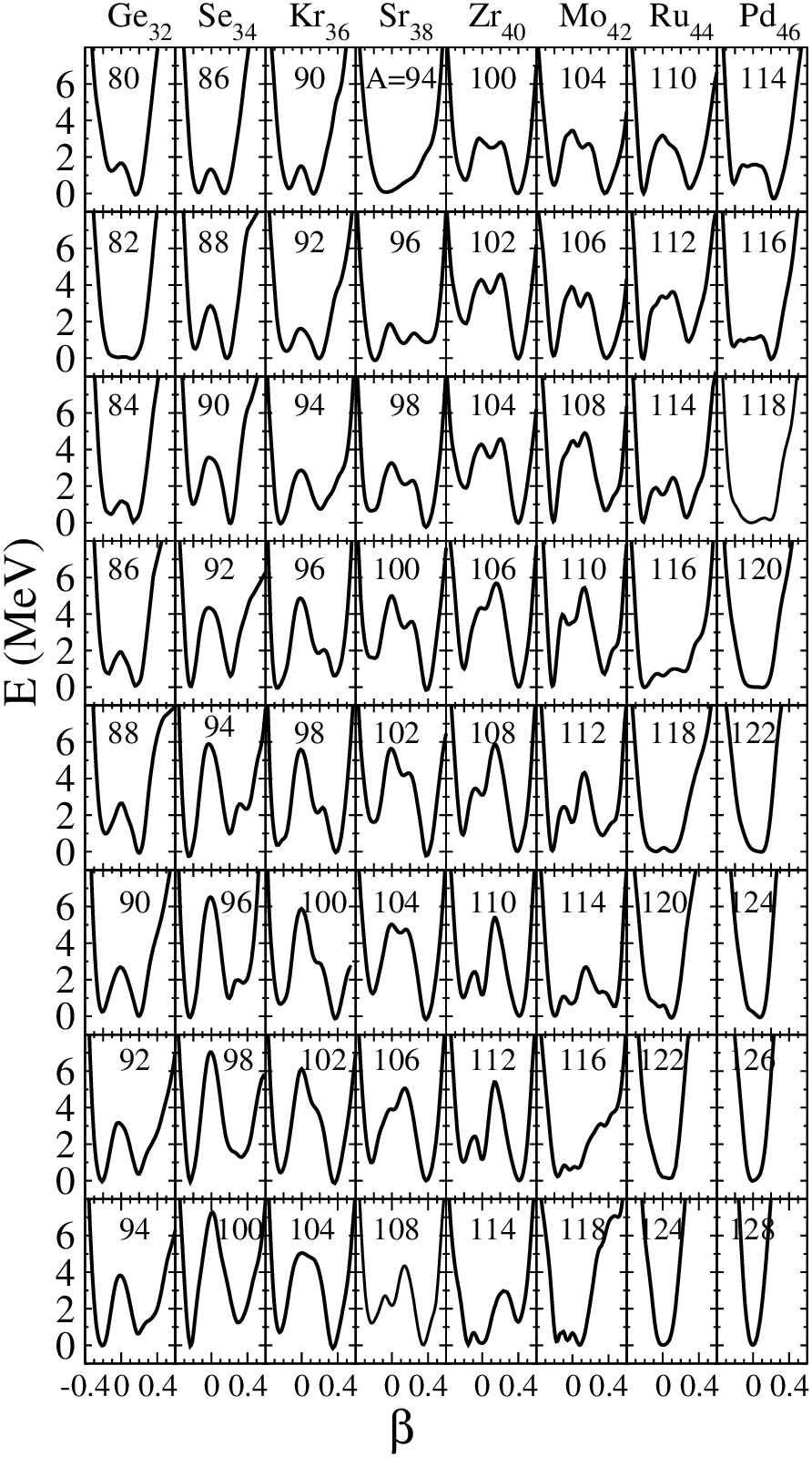}
\caption{Potential energy curves for even-even neutron-rich Ge, Se, Kr, 
Sr, Zr, Mo, Ru, and Pd isotopes obtained from constrained HF+BCS 
calculations with the Skyrme force SLy4.}
\label{fig_eq}
\end{figure}

\begin{figure}[ht]
\centering
\includegraphics[width=85mm]{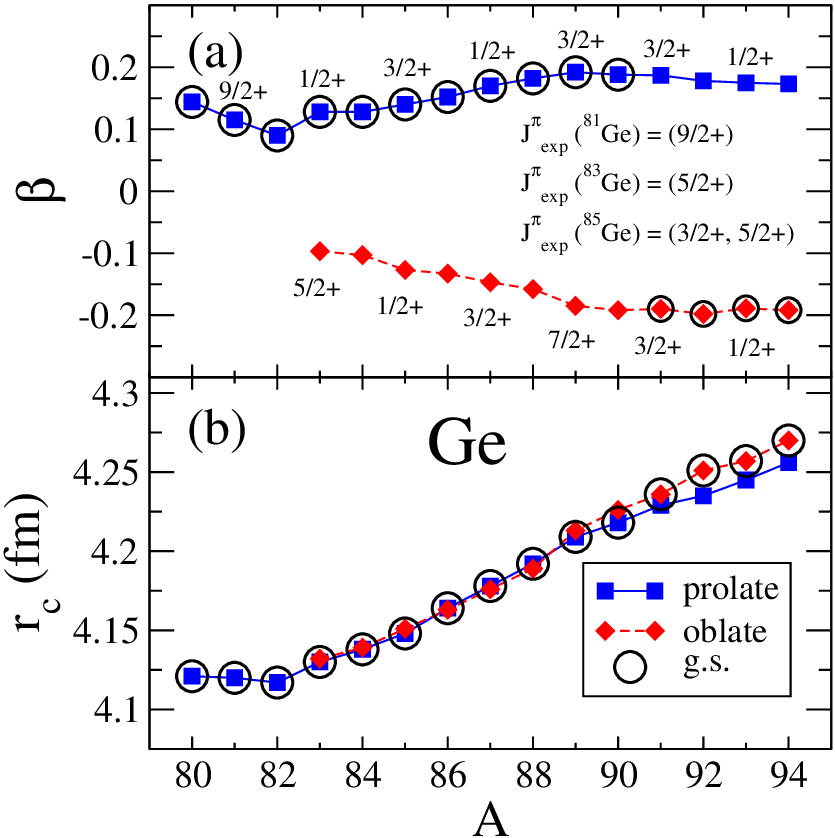}
\caption{(Color online) Isotopic evolution of the quadrupole deformation parameter 
$\beta$ (a) and charge radius (b) corresponding to the energy minima obtained from
the Skyrme interaction SLy4 for Ge isotopes. Ground state results are encircled.}
\label{ge_beta_rc}
\end{figure}

\begin{figure}[ht]
\centering
\includegraphics[width=85mm]{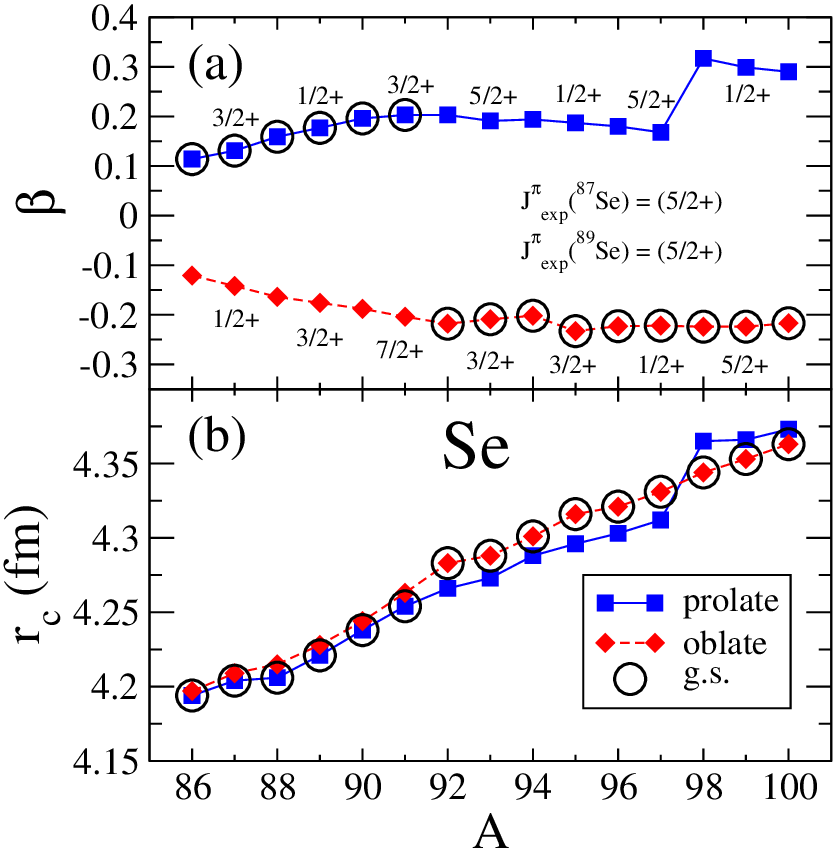}
\caption{(Color online) Same as in  Fig. \ref{ge_beta_rc}, but for Se
isotopes.}
\label{se_beta_rc}
\end{figure}

\begin{figure}[ht]
\centering
\includegraphics[width=85mm]{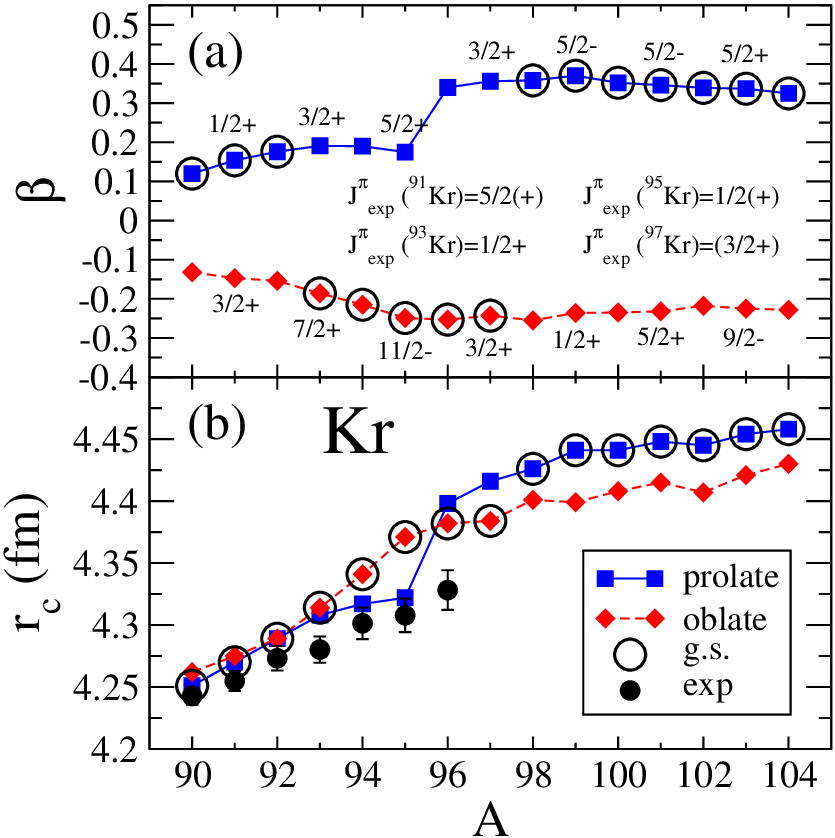}
\caption{(Color online) Same as in  Fig. \ref{ge_beta_rc}, but for Kr
isotopes. Experimental charge radii are from \cite{angeli04}.}
\label{kr_beta_rc}
\end{figure}

\begin{figure}[ht]
\centering
\includegraphics[width=85mm]{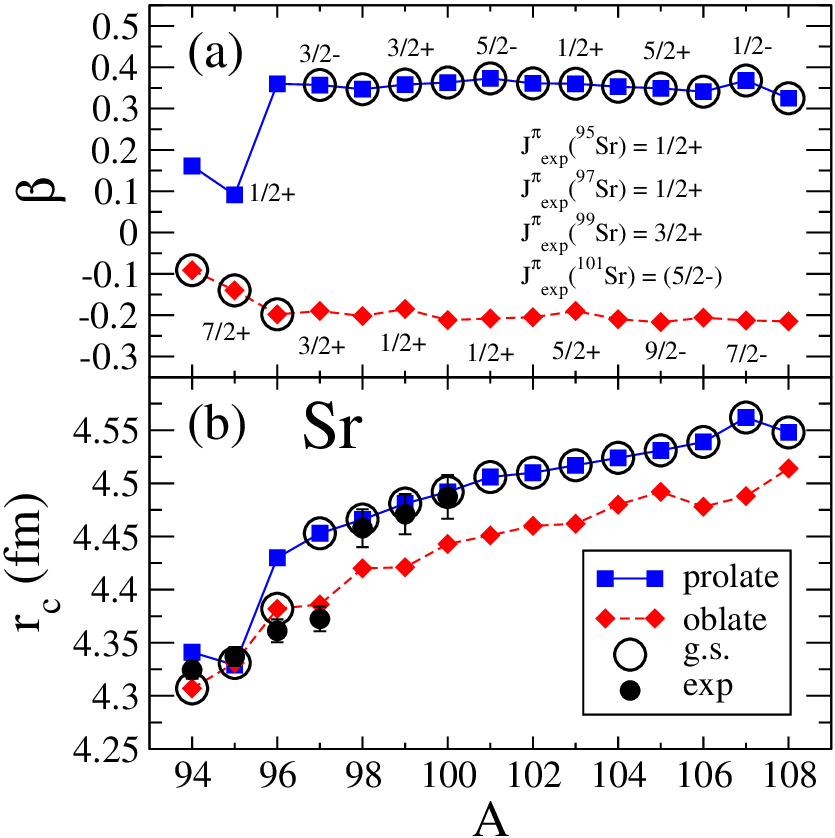}
\caption{(Color online) Same as in  Fig. \ref{ge_beta_rc}, but for Sr
isotopes. Experimental charge radii are from \cite{angeli04}.}
\label{sr_beta_rc}
\end{figure}

\begin{figure}[ht]
\centering
\includegraphics[width=85mm]{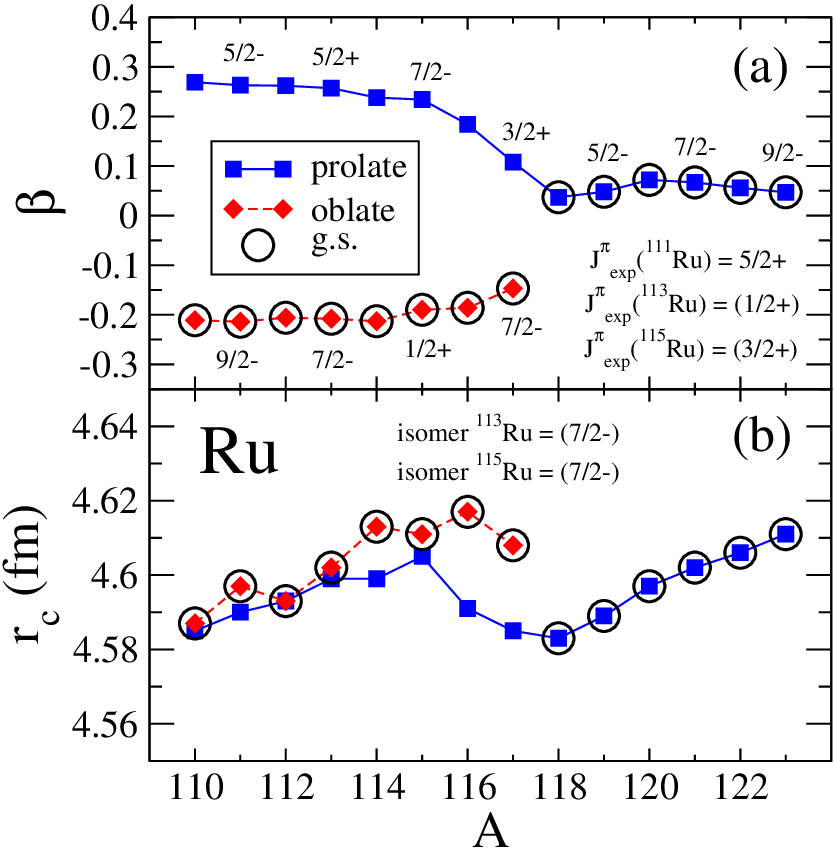}
\caption{(Color online) Same as in  Fig. \ref{ge_beta_rc}, but for Ru
isotopes.}
\label{ru_beta_rc}
\end{figure}

\begin{figure}[ht]
\centering
\includegraphics[width=85mm]{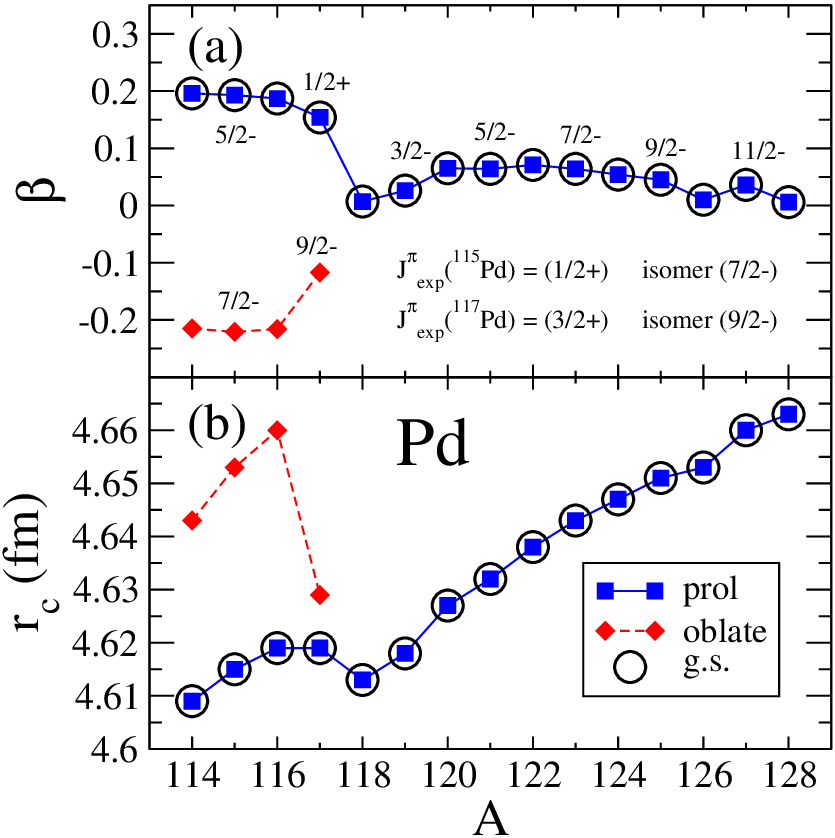}
\caption{(Color online) Same as in  Fig. \ref{ge_beta_rc}, but for Pd
isotopes.}
\label{pd_beta_rc}
\end{figure}

\begin{figure}[ht]
\centering
\includegraphics[width=70mm]{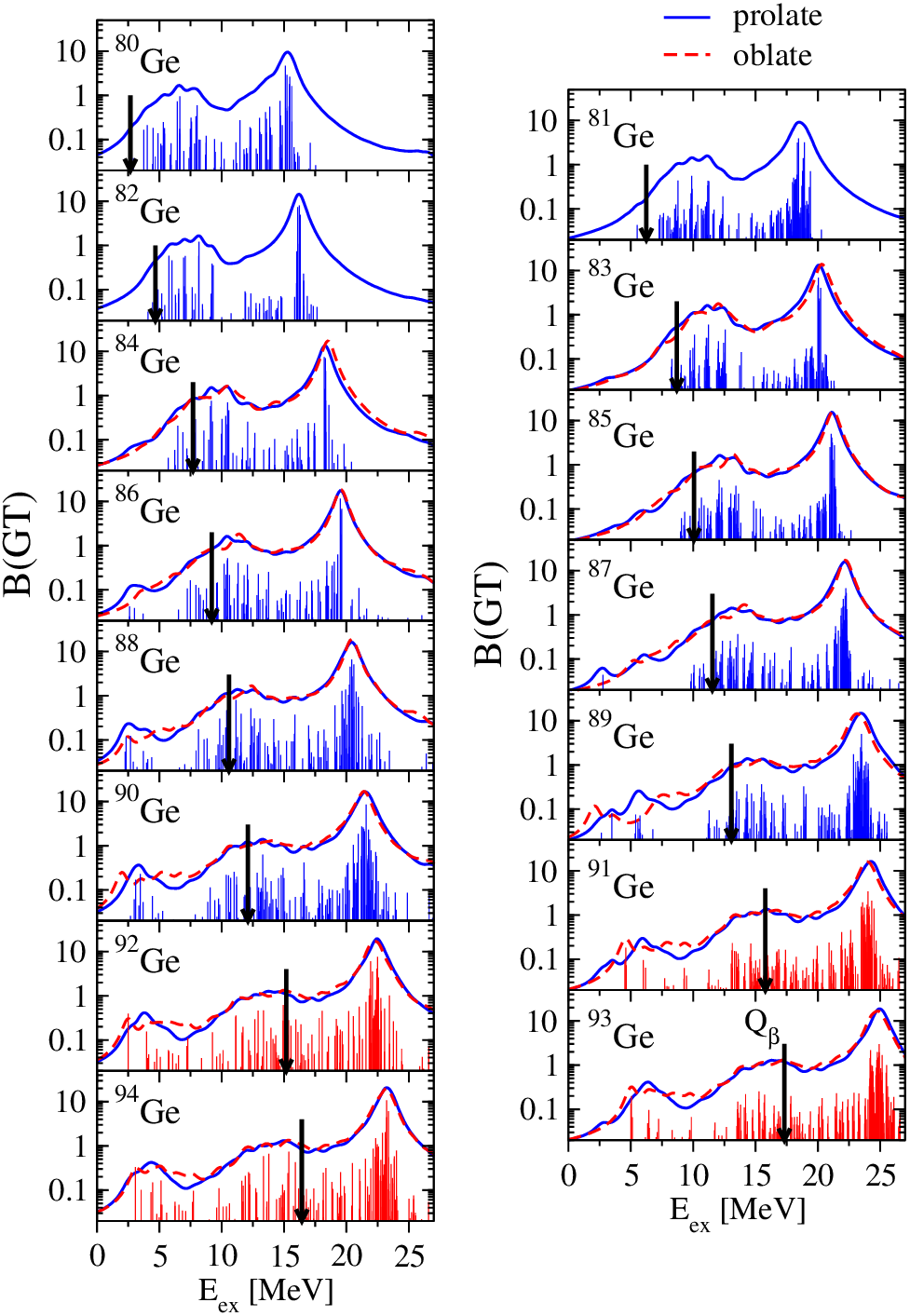}
\caption{(Color online) QRPA-SLy4 Gamow-Teller strength distributions 
for Ge isotopes as a function of the excitation energy in the daughter
nucleus. The calculations correspond to the various equilibrium
deformations found in the PECs.}
\label{fig_ge_fold}
\end{figure}

\begin{figure}[ht]
\centering
\includegraphics[width=70mm]{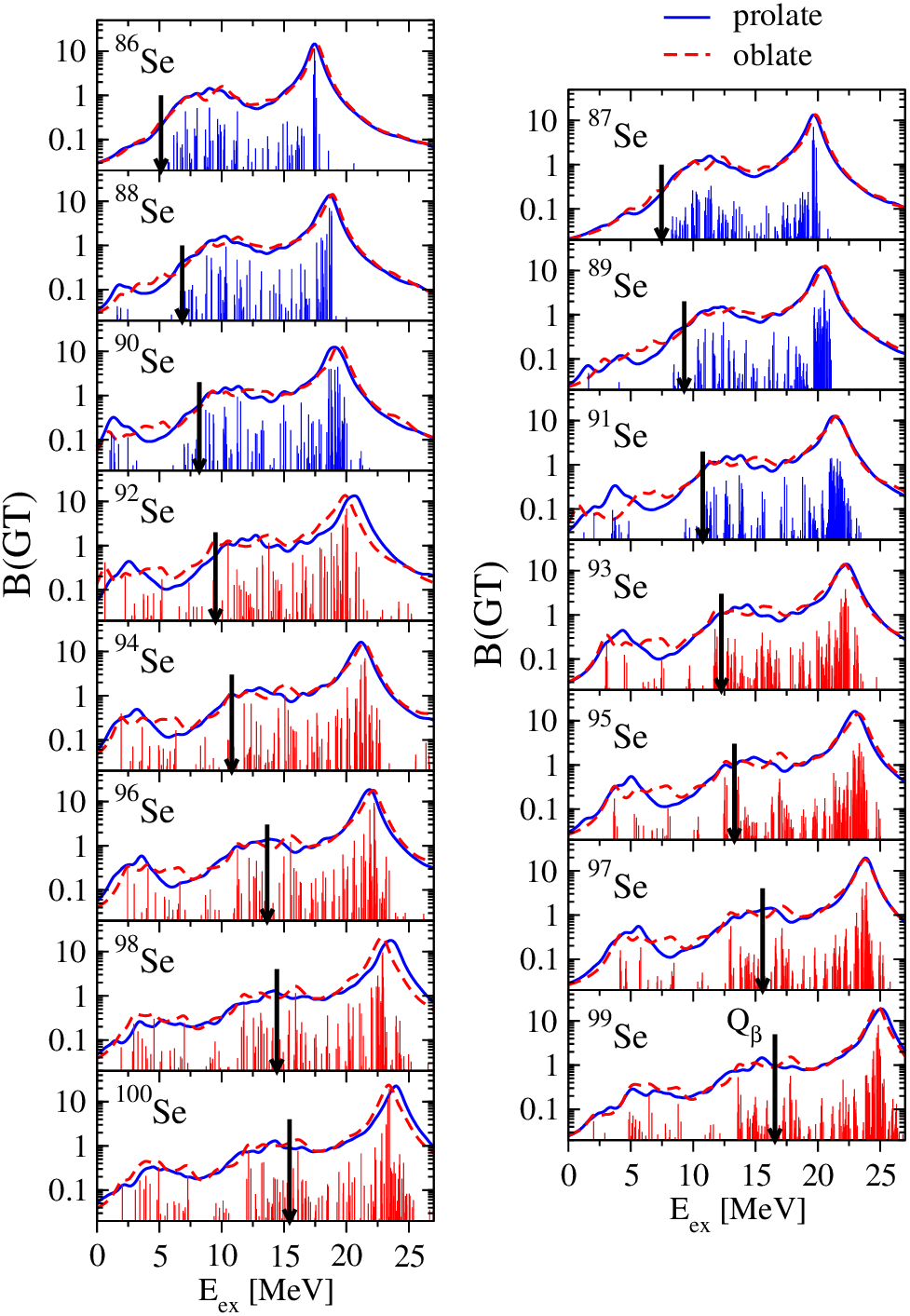}
\caption{(Color online) Same as in  Fig. \ref{fig_ge_fold}, but for Se 
isotopes.}
\label{fig_se_fold}
\end{figure}

\begin{figure}[ht]
\centering
\includegraphics[width=70mm]{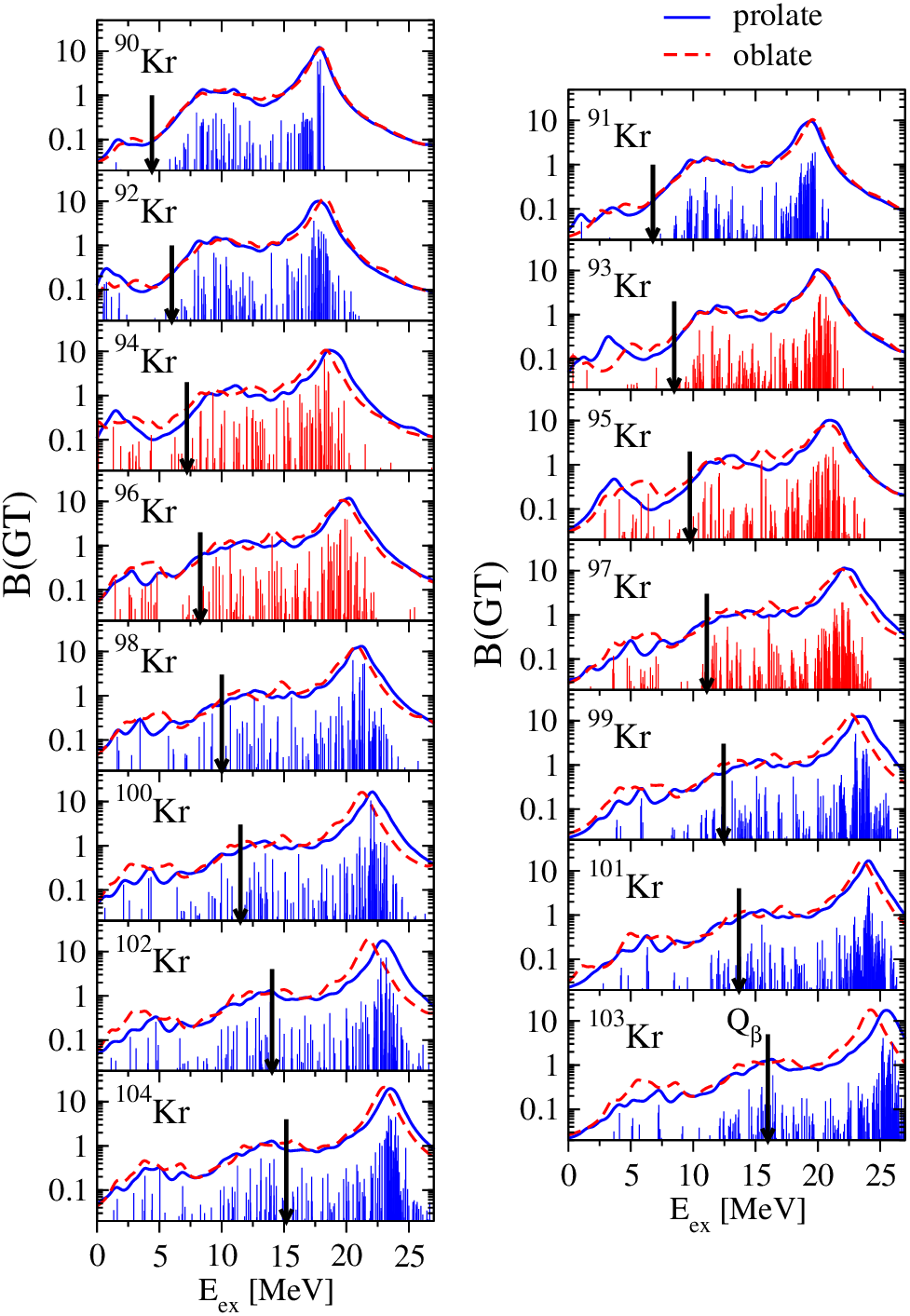}
\caption{(Color online) Same as in  Fig. \ref{fig_ge_fold}, but for Kr 
isotopes.}
\label{fig_kr_fold}
\end{figure}

\begin{figure}[ht]
\centering
\includegraphics[width=70mm]{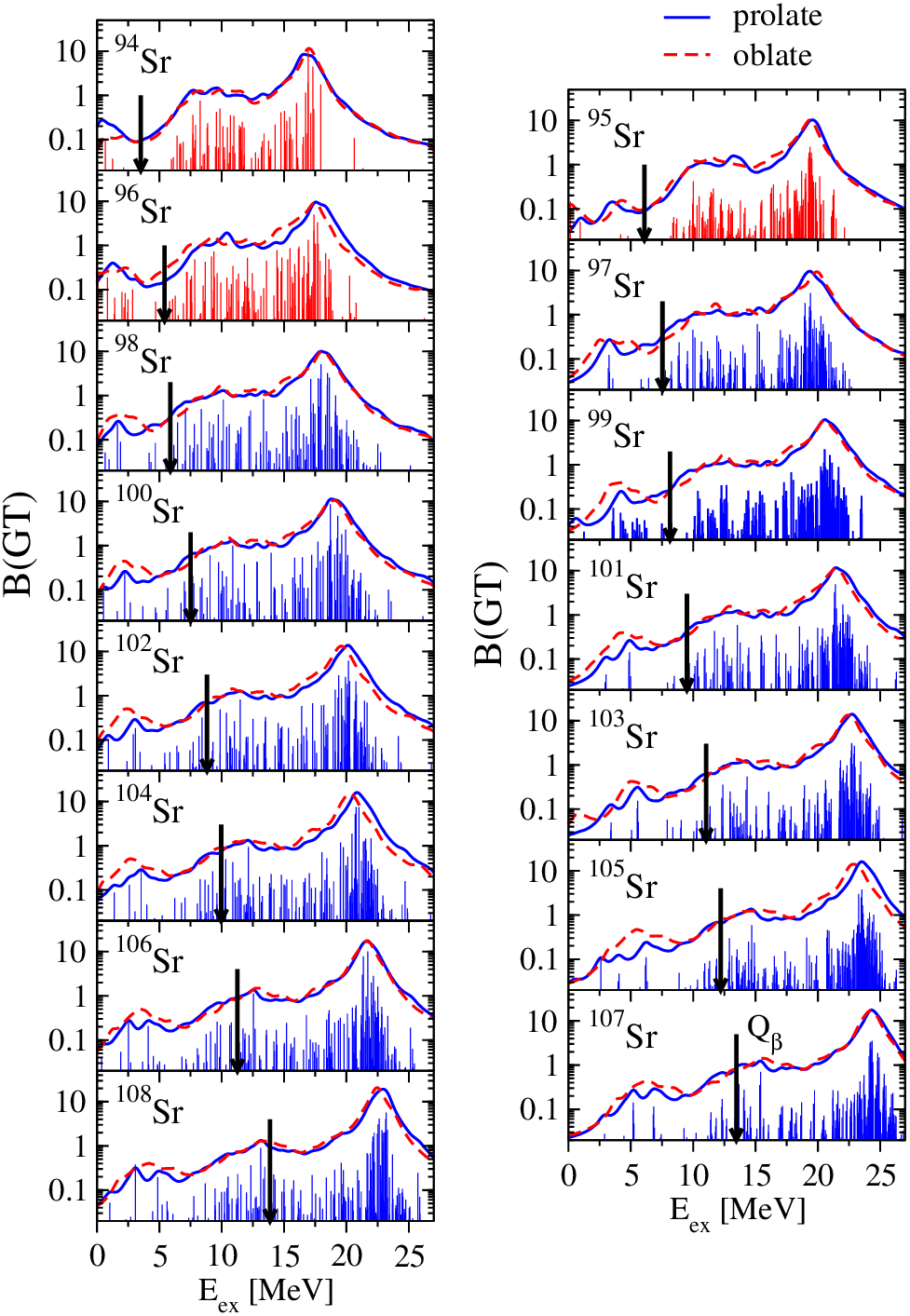}
\caption{(Color online) Same as in  Fig. \ref{fig_ge_fold}, but for Sr
isotopes.}
\label{fig_sr_fold}
\end{figure}

\begin{figure}[ht]
\centering
\includegraphics[width=70mm]{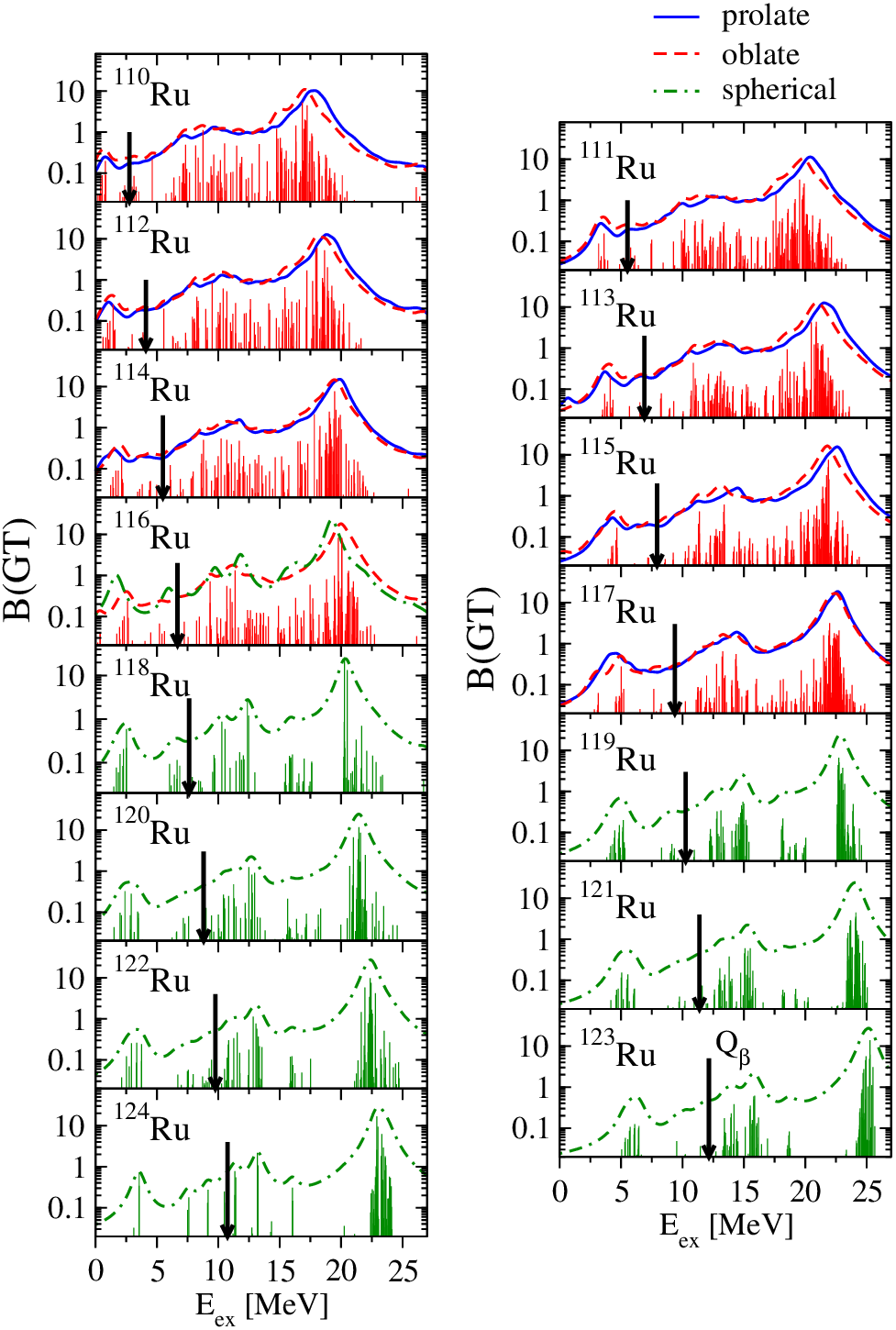}
\caption{(Color online) Same as in  Fig. \ref{fig_ge_fold}, but for Ru 
isotopes.}
\label{fig_ru_fold}
\end{figure}

\begin{figure}[ht]
\centering
\includegraphics[width=70mm]{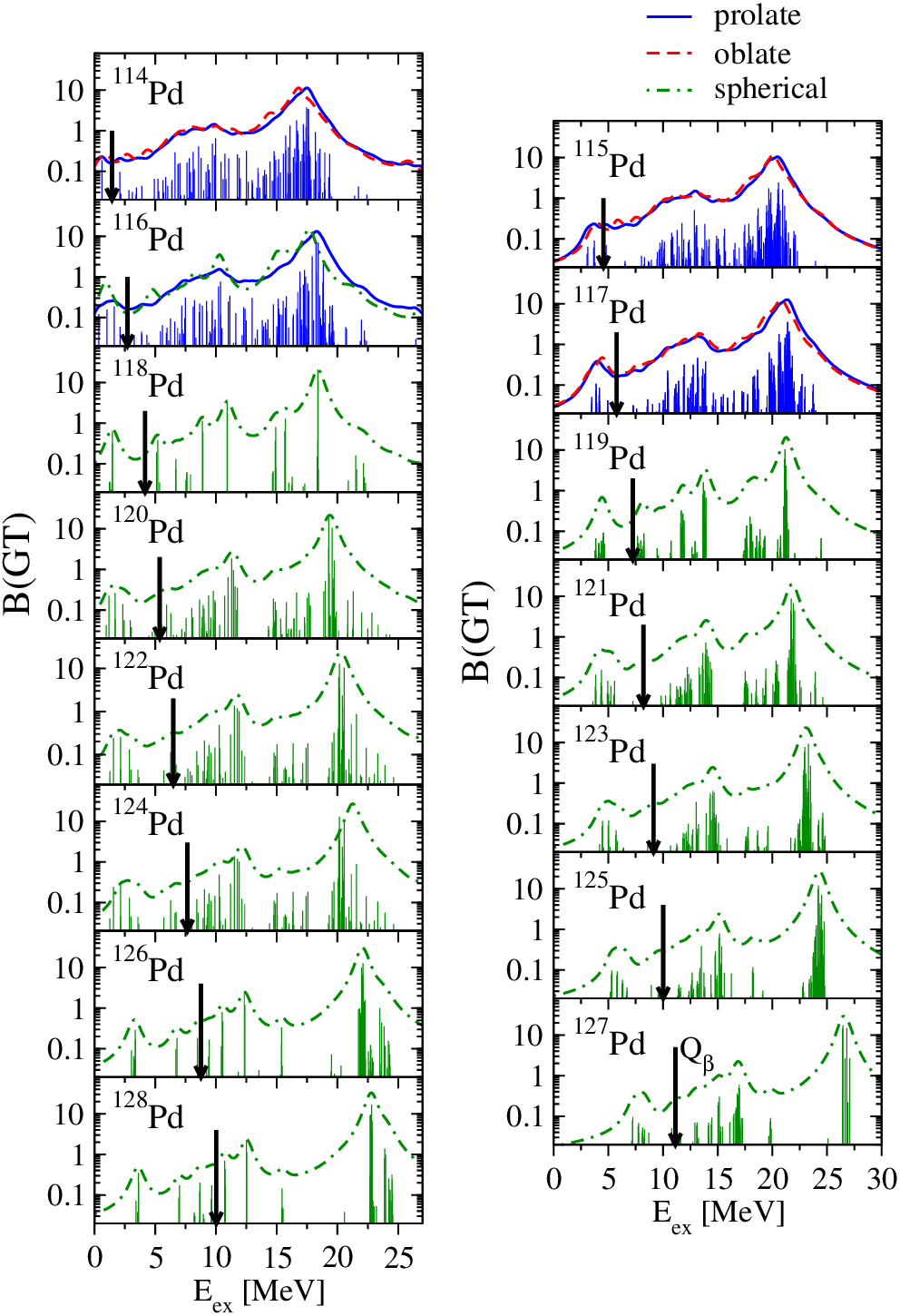}
\caption{(Color online)  Same as in  Fig. \ref{fig_ge_fold}, but for Pd 
isotopes.}
\label{fig_pd_fold}
\end{figure}

\clearpage

The $\beta$-decay half-life is obtained by summing all the allowed
transition strengths to states in the daughter nucleus with
excitation energies lying below the corresponding $Q$-energy,
$Q_\beta\equiv Q_{\beta^-}= M(A,Z)-M(A,Z+1)-m_e $, written in terms of 
the nuclear masses $M(A,Z)$ and the electron mass ($m_e$), and
weighted with the phase space factors $f(Z,Q_{\beta}-E_{ex})$,

\begin{equation}
T_{1/2}^{-1}=\frac{\left( g_{A}/g_{V}\right) _{\rm eff} ^{2}}{D}
\sum_{0 < E_{ex} < Q_\beta}f\left( Z,Q_{\beta}-E_{ex} \right) B(GT,E_{ex}) \, ,
 \label{t12}
\end{equation}
with $D=6200$~s and $(g_A/g_V)_{\rm eff}=0.77(g_A/g_V)_{\rm free}$,
where 0.77 is a standard quenching factor. The same quenching factor
is included in all the figures shown later for the GT strength
distributions. The bare results can be 
recovered by scaling the results in this paper for $B(GT)$ and 
$T_{1/2}$ with the square of this quenching factor.

The Fermi integral $f(Z,Q_{\beta}-E_{ex})$ is computed numerically for 
each value of the energy including screening and finite size effects, 
as explained in Ref. \cite{gove},

\begin{equation}
f^{\beta^\pm} (Z, W_0) = \int^{W_0}_1 p W (W_0 - W)^2 \lambda^\pm(Z,W)
{\rm d}W\, ,
\end{equation}
with

\begin{equation}
\lambda^\pm(Z,W) = 2(1+\gamma) (2pR)^{-2(1-\gamma)} e^{\mp\pi y}
\frac{|\Gamma (\gamma+iy)|^2}{[\Gamma (2\gamma+1)]^2}\, ,
\end{equation}
where $\gamma=\sqrt{1-(\alpha Z)^2}$ ; $y=\alpha ZW/p$ ; $\alpha$ is
the fine structure constant and $R$ the nuclear radius. $W$ is the
total energy of the $\beta$ particle, $W_0$ is the total energy
available in $m_e c^2$ units, and $p=\sqrt{W^2 -1}$ is the momentum
in $m_e c$ units.

\section{Results and discussion}
\label{results}

In this section I present first the PECs in the isotopic chains studied.
Quadrupole deformation parameters as well as charge 
r.m.s. radii ($r_c$) are analyzed as a function of the mass number.
Then, energy distributions of the GT strength corresponding to the local 
minima in the PECs are calculated. 
Finally, half-lives are evaluated and compared with the experiment.

\subsection{Structural isotopic evolution}

In Fig. \ref{fig_eq} the PECs, i.e., the energies relative 
to that of the ground state, are plotted as a function of the quadrupole 
deformation $\beta$ for the neutron-rich Ge, Se, Kr, Sr, Zr, Mo, 
Ru, and Pd isotopes. The results correspond to the SLy4 interaction.
The isotopes covered in this study include middle-shell nuclei with
proton numbers between shell closures $Z=28$ and $Z=50$, namely 
$Z=32,34,36,38,40,42,44,46$ and neutron numbers between shell 
closures $N=50$ (as in $^{82}$Ge) and $N=82$ (as in the 
heaviest $^{128}$Pd).

In most of the isotopic chains one can see the appearance of several
equilibrium nuclear shapes, whose relative energies change with the 
number of neutrons. In Ge isotopes, prolate shapes that are
ground states in the lighter isotopes are found with the only exception of  
$^{82}$Ge, where a spherical shape is found in accordance with its
$N=50$ semi-magic character. At $N=58,60$ ($^{90,92}$Ge) oblate and 
prolate shapes are practically degenerate in energy and oblate 
shapes become ground states for heavier isotopes. The case of Se
isotopes is similar with oblate and prolate minima all along
the isotopic chain. The lighter (heavier) isotopes have prolate 
(oblate) ground states with transitional isotopes around $N=58,60$ 
($^{92,94}$Se). In this case the energy barriers are more pronounced 
than in the case of Ge isotopes. 
Kr isotopes show competing shapes in the lighter isotopes that
become oblate at $N=58,60$ ($^{94,96}$Kr) and then turn into prolate
shapes beyond $^{98}$Kr. In the heavier isotopes, as in the case of
Se isotopes,  shape coexistence is found with very well developed 
oblate and prolate minima 
separated with high energy barriers. Sr isotopes show a transition
from oblate at $N=58$ ($^{96}$Sr) to prolate at $N=60$ ($^{98}$Sr)
with a two minima structure for heavier isotopes. 
The cases of Zr and Mo isotopes were discussed in 
Refs. \cite{sarripere,sarri14}. Both oblate and prolate minima
are observed in the lighter isotopes of Zr and Mo with prolate
ground states. Whereas the prolate shape remains ground state in
most of the heavier Zr isotopes, oblate shapes are lower in energy 
for the heavier Mo isotopes. Finally, Ru and Pd isotopes show oblate
and prolate minima in the lighter isotopes and a gradual transition 
into spherical shapes as one approaches the shell closure at $N=82$.

In summary, a large diversity of nuclear structures are found in this 
mass region, from spherical to well deformed shapes, passing
through soft transitional nuclei and even possible shape-coexistence 
structures. This rich variety of shapes represents a challenge to any
theoretical model trying to describe them in a unified manner. In the 
next subsections the results are compared with the available experimental 
data, which are restricted at present to $\beta$-decay half-lives.
Then, the theoretical approach will be tested against this information
and the limitations of the model will be established.

These results are in qualitative agreement with similar calculations
obtained in this mass region from different theoretical approaches, 
including macroscopic-microscopic methods based on liquid drop models 
with shell corrections \cite{skalski97,FRDM}, relativistic mean 
fields \cite{lala2}, as well as nonrelativistic calculations with 
Skyrme \cite{bonche85} and Gogny  \cite{rayner1,rayner2,hilaire} 
interactions. Thus, a consistent theoretical description emerges, 
which is supported by the still scarce experimental information 
available 
\cite{heyde11,wu04,urban04,sumikama11,mach,goodin,urban,cheal10,albers}.

The isotopic evolution can be better appreciated in Figs. 
\ref{ge_beta_rc}--\ref{pd_beta_rc},
where quadrupole deformations $\beta$ (a) and r.m.s. charge
radii $r_c$ (b) of the various energy minima are plotted
as a function of the mass number $A$. The deformation corresponding 
to the ground state for each isotope is encircled.  Also shown in 
these figures for odd-$A$ isotopes, are the spin and parity $(J^{\pi})$ 
of the different shapes and the experimental assignments \cite{audi12}. 
The experimental assignments based on systematics estimated from 
trends in neighboring nuclides have not been included.

In Fig. \ref{ge_beta_rc} for Ge isotopes one can see clearly the 
shape transition at $A=90-92\ (N=58-60)$ from prolate shapes with 
$\beta\approx 0.2$ to oblate shapes  with $\beta\approx -0.2$. 
Charge r.m.s. radii have not been measured in these isotopes, but 
it is expected from these calculations a very smooth behavior given 
that the magnitude of $\beta$ in the prolate and oblate sectors are 
very similar. Fig. \ref{se_beta_rc} shows the analogous results for 
Se isotopes. In this case one can see the transition from prolate 
($\beta\approx 0.2$) to oblate ($\beta\approx -0.2$) at 
$A=92\ (N=58)$. The prolate shape grows in the heavier isotopes 
($\beta\approx 0.3$), but they are never ground states and then, the 
expected increasing in the charge radii is smooth.  
Kr isotopes in Fig. \ref{kr_beta_rc} show first a shape transition
from prolate ($\beta\approx 0.15$) to oblate ($\beta\approx -0.25$) 
and a subsequent transition from oblate to prolate  ($\beta\approx 0.35$) 
shapes. The radii are sensitive to this transitions, although the measured 
radii \cite{angeli04} seem to favored prolate shapes in the lighter 
isotopes. Sr isotopes in Fig. \ref{sr_beta_rc} show a clear transition 
from oblate to strong prolate ($\beta\approx 0.4$) deformations at 
$A=96-98\ (N=58-60)$. This shape transition is well correlated with the 
trend change observed in the charge radii that shows a sizable jump 
between $^{96}$Sr and $^{98}$Sr both theoretical and 
experimentally \cite{angeli04}.
In the case of Ru (Pd) isotopes shown in Fig. \ref{ru_beta_rc} 
(\ref{pd_beta_rc}), one can see a smooth transition from deformed oblate 
(prolate) solutions in the lighter isotopes to spherical shapes in the 
heavier ones. This change is felt in the trends of the radii, but no 
experiments are yet available to compare with.

Spins and parities in odd-$A$ isotopes can be compared with their
experimental assignments. In the Ge isotopes the calculations agree
reasonably well with the assignments taking into account that oblate 
and prolate shapes are very close in energy and that a $1/2^+$ isomer 
is observed experimentally in $^{83}$Ge at 248 keV. In the lighter 
Se isotopes, $1/2^+$ and $3/2^+$ states are obtained, whereas experimental 
assignments are $(5/2^+)$. In both isotopes, $5/2^+$ states 
very close in energy to the ground states are also obtained, although 
somewhat above. Similarly, in the lighter Kr isotopes the experimental 
assignments are obtained 
very close in energy to the ground states, although slightly above. On 
the other hand, a $7/2^+$ isomer is experimentally observed in $^{93}$Kr
at 355 keV that corresponds to the ground state here. Sr isotopes exhibit 
a nice agreement. The measured spin and parities of ground states in 
$^{95,99,101}$Sr correspond to the prolate calculations. A $(7/2^+)$ state 
is also observed experimentally in $^{95}$Sr at 56 keV. In $^{97}$Sr  
the observed $1/2^+$ ground state appears as an excited state. It is also 
worth noting that the prolate ground state $(3/2^-)$ for this isotope 
is observed experimentally at 645 keV. In the case of Ru isotopes  
the measured $J^{\pi}$ are difficult to reproduce. They are found in the 
calculations, but not as ground states. On the other 
hand, the negative parity $7/2^-$ states found in the calculations 
are also seen experimentally at low energies. In particular, an 
isomeric state $(7/2^-)$ at an undetermined energy has been seen 
in $^{113}$Ru. Finally, in Pd isotopes the negative parity isomers,
which are oblate in this description, are reproduced in the calculations, but not
the ground states.

\begin{figure}[ht]
\centering
\includegraphics[width=85mm]{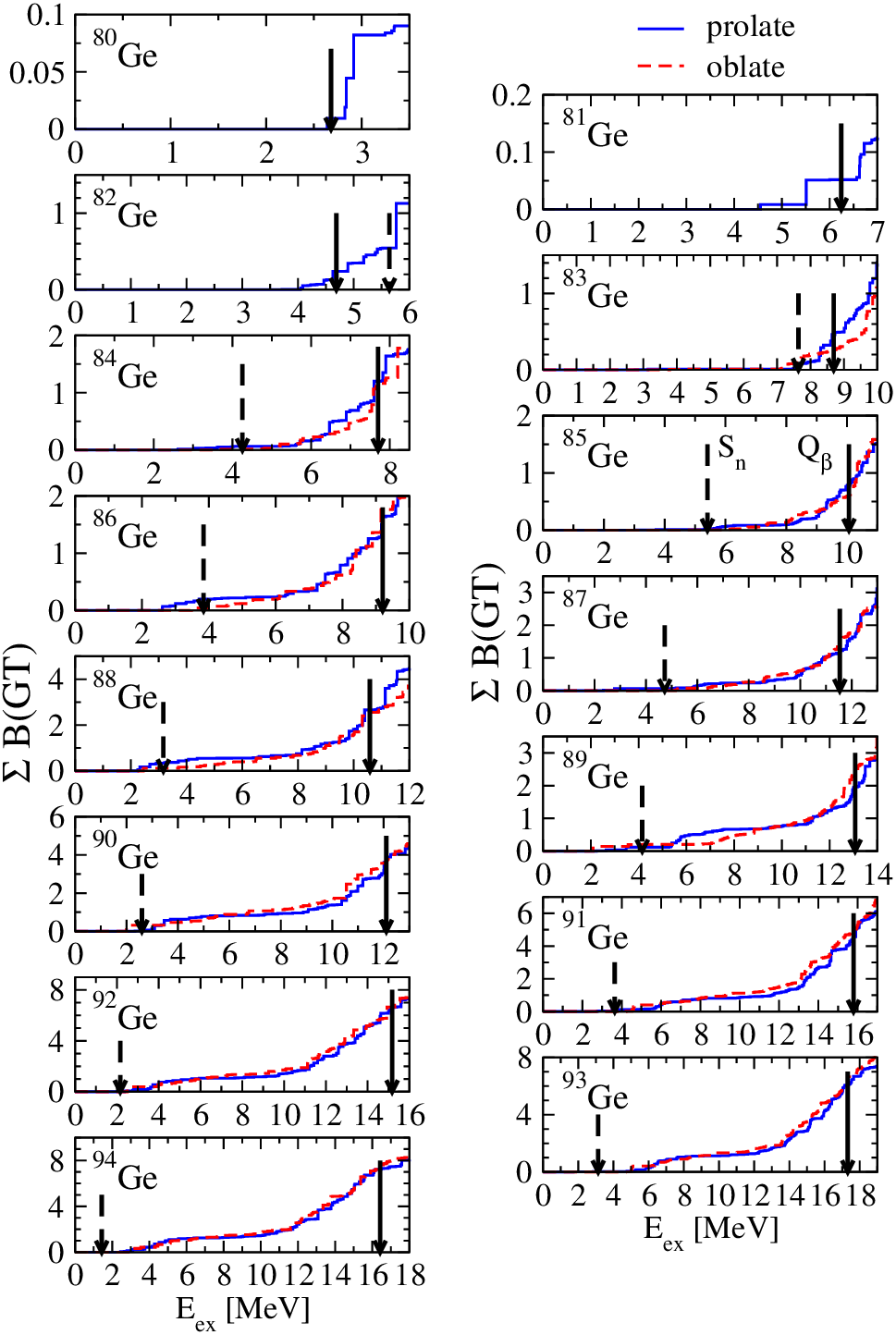}
\caption{(Color online) QRPA-SLy4 accumulated GT strengths in Ge isotopes 
calculated for the various equilibrium shapes. 
$Q_\beta$ and $S_n$ energies are shown by solid and dashed vertical 
arrows, respectively.}
\label{fig_ge_acum}
\end{figure}

\begin{figure}[ht]
\centering
\includegraphics[width=85mm]{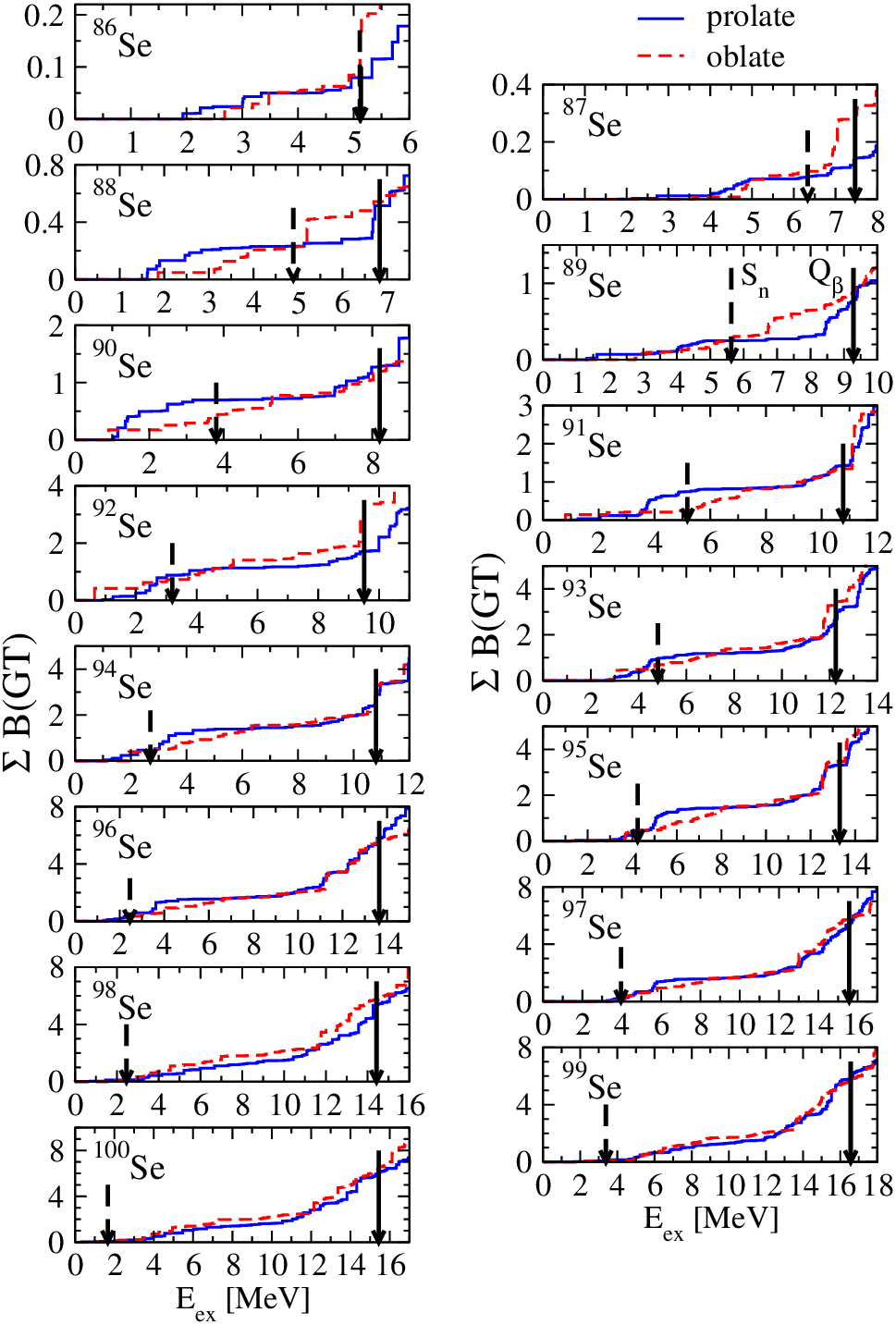}
\caption{(Color online)  Same as in  Fig. \ref{fig_ge_acum}, but for Se
isotopes.}
\label{fig_se_acum}
\end{figure}

\begin{figure}[ht]
\centering
\includegraphics[width=85mm]{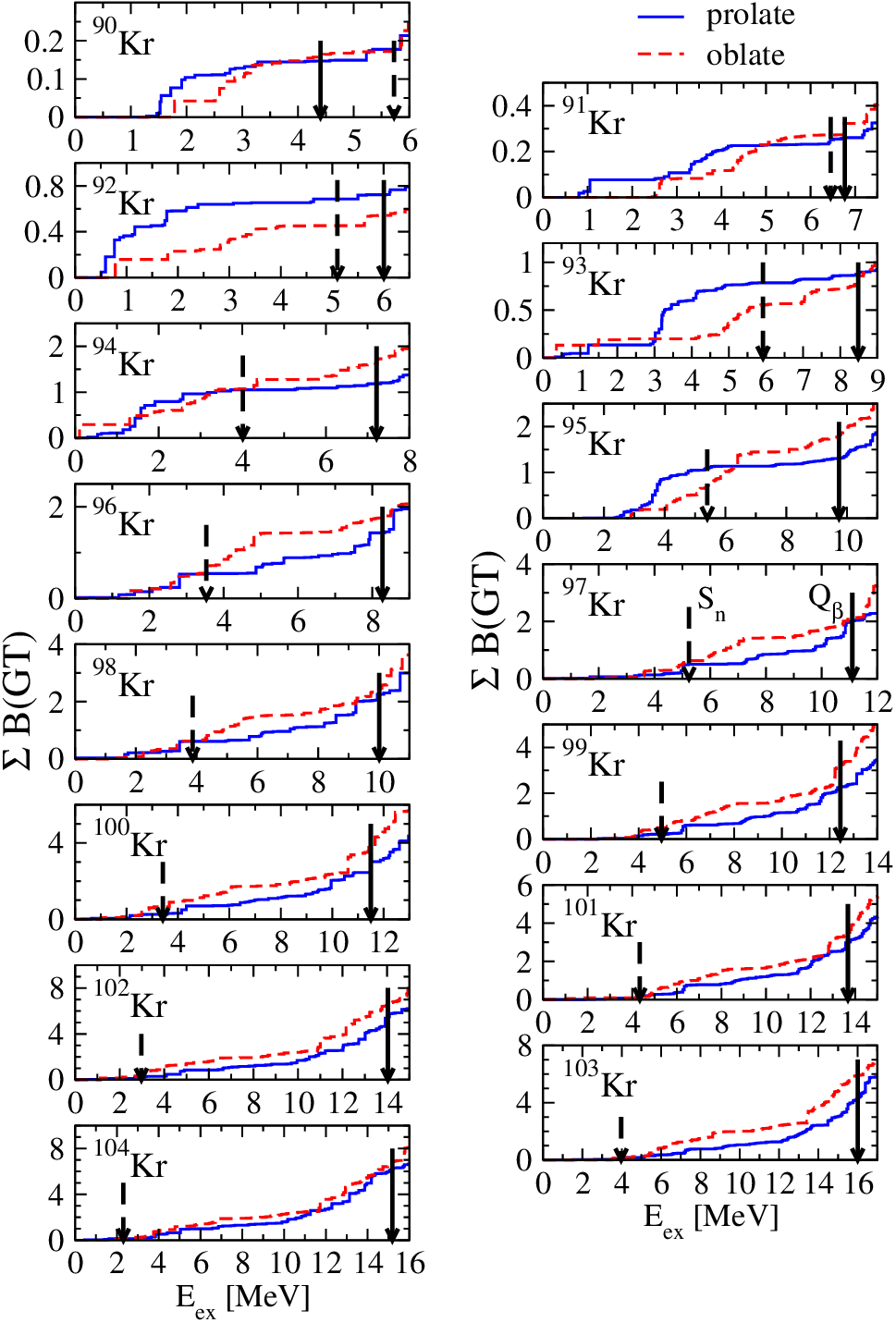}
\caption{(Color online) Same as in  Fig. \ref{fig_ge_acum}, but for Kr
isotopes.}
\label{fig_kr_acum}
\end{figure}

\begin{figure}[ht]
\centering
\includegraphics[width=85mm]{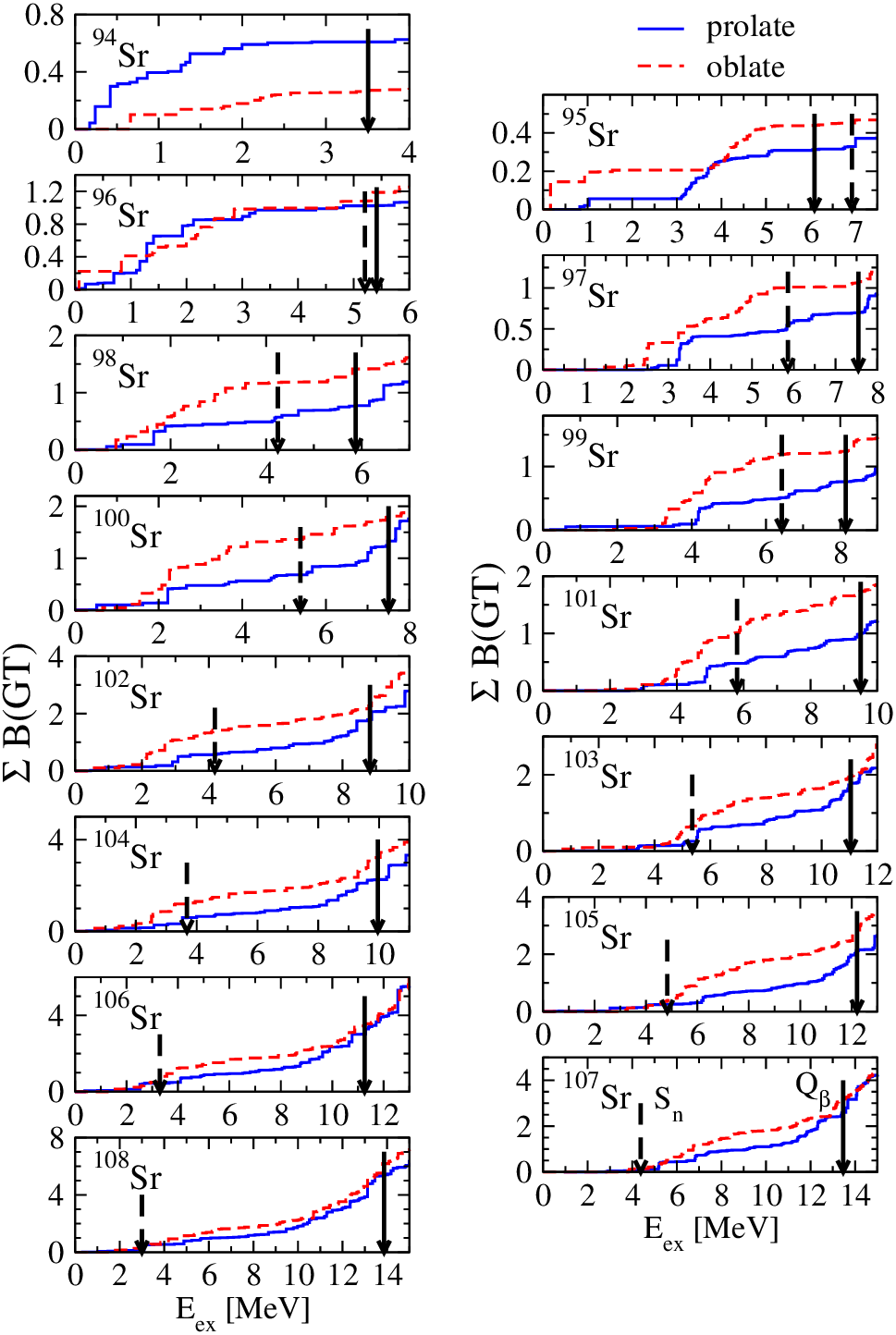}
\caption{(Color online) Same as in  Fig. \ref{fig_ge_acum}, but for Sr
isotopes.}
\label{fig_sr_acum}
\end{figure}

\begin{figure}[ht]
\centering
\includegraphics[width=85mm]{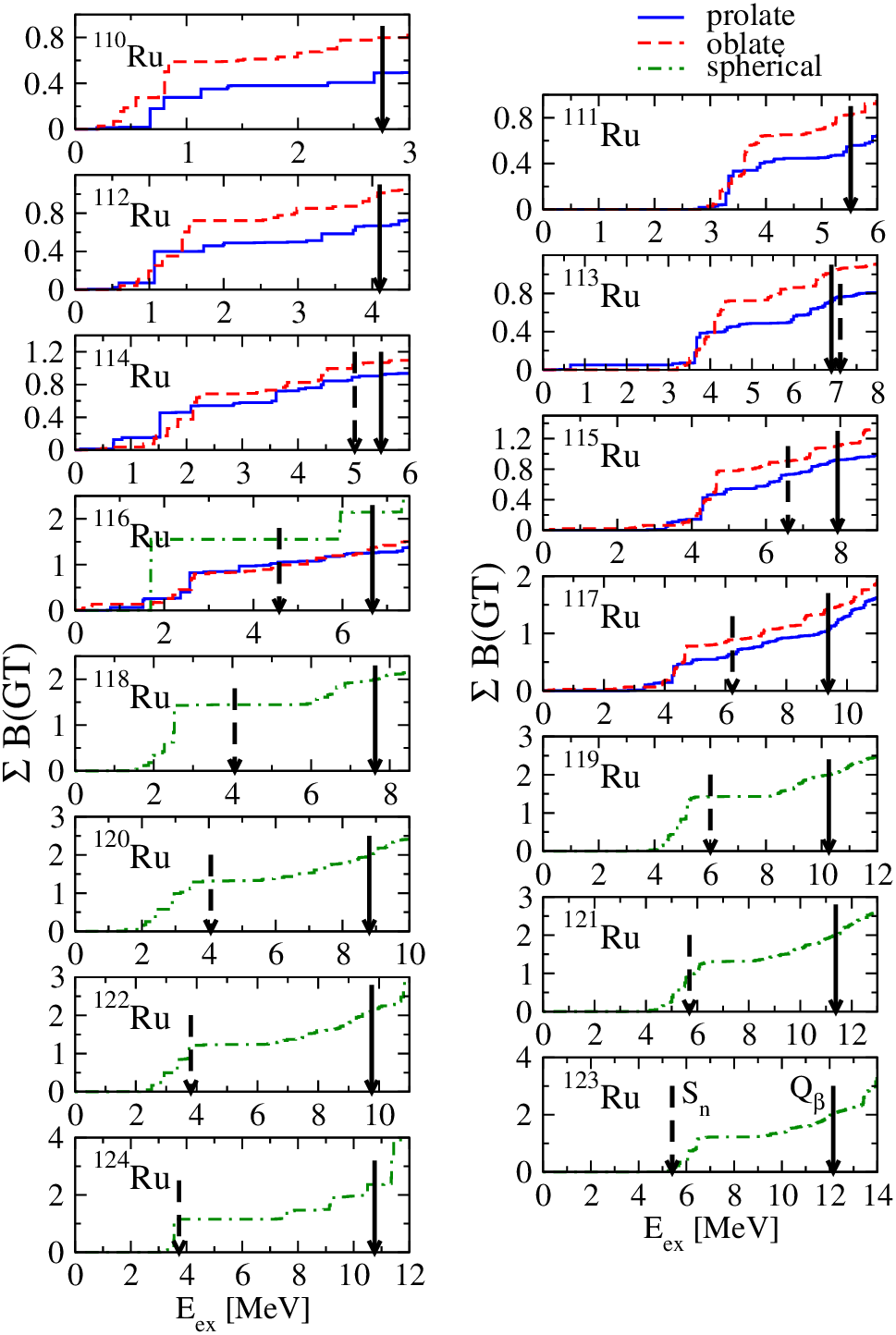}
\caption{(Color online) Same as in  Fig. \ref{fig_ge_acum}, but for Ru
isotopes.}
\label{fig_ru_acum}
\end{figure}

\begin{figure}[ht]
\centering
\includegraphics[width=85mm]{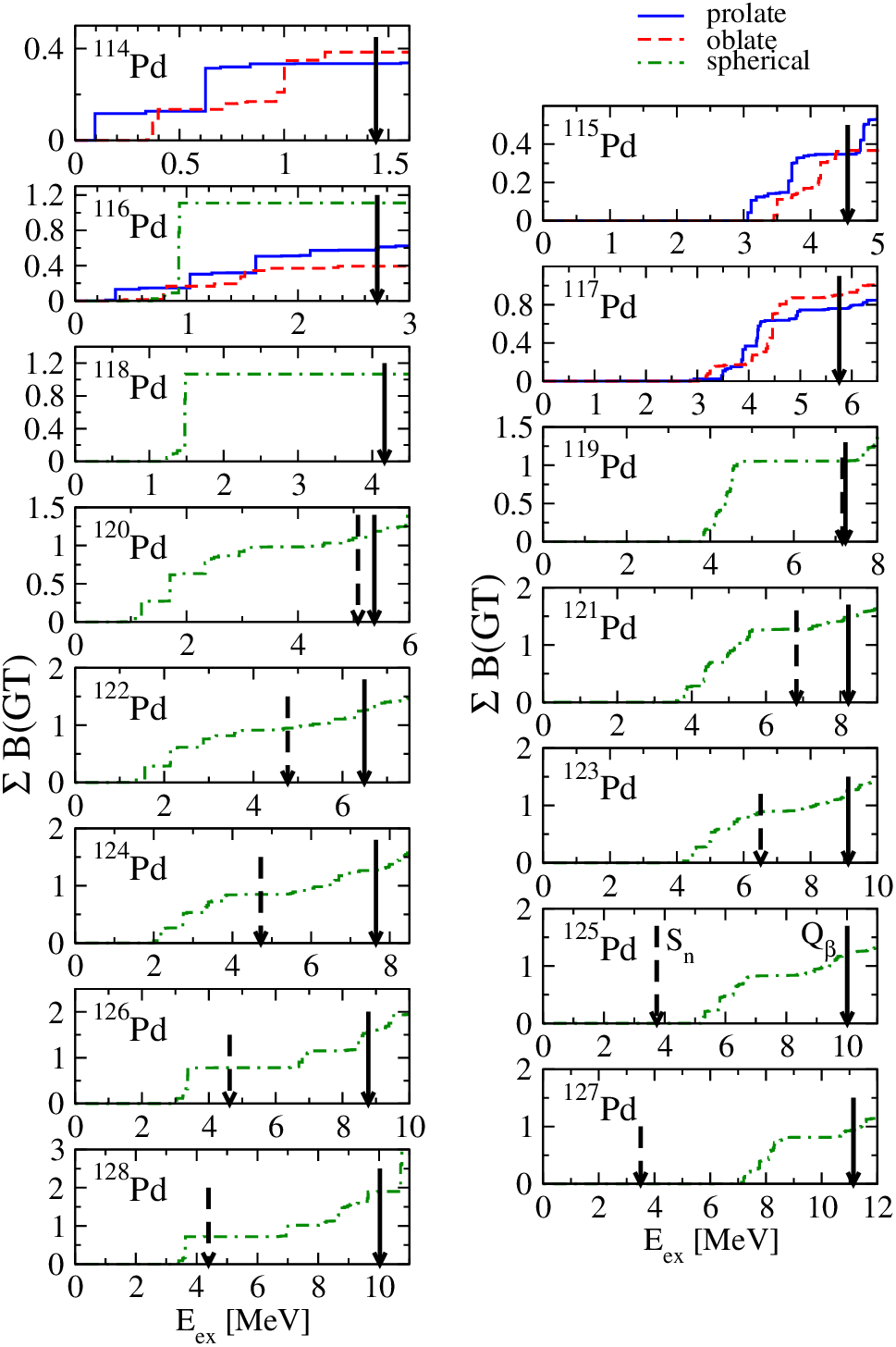}
\caption{(Color online) Same as in  Fig. \ref{fig_ge_acum}, but for Pd
isotopes.}
\label{fig_pd_acum}
\end{figure}

\begin{figure}[ht]
\centering
\includegraphics[width=85mm]{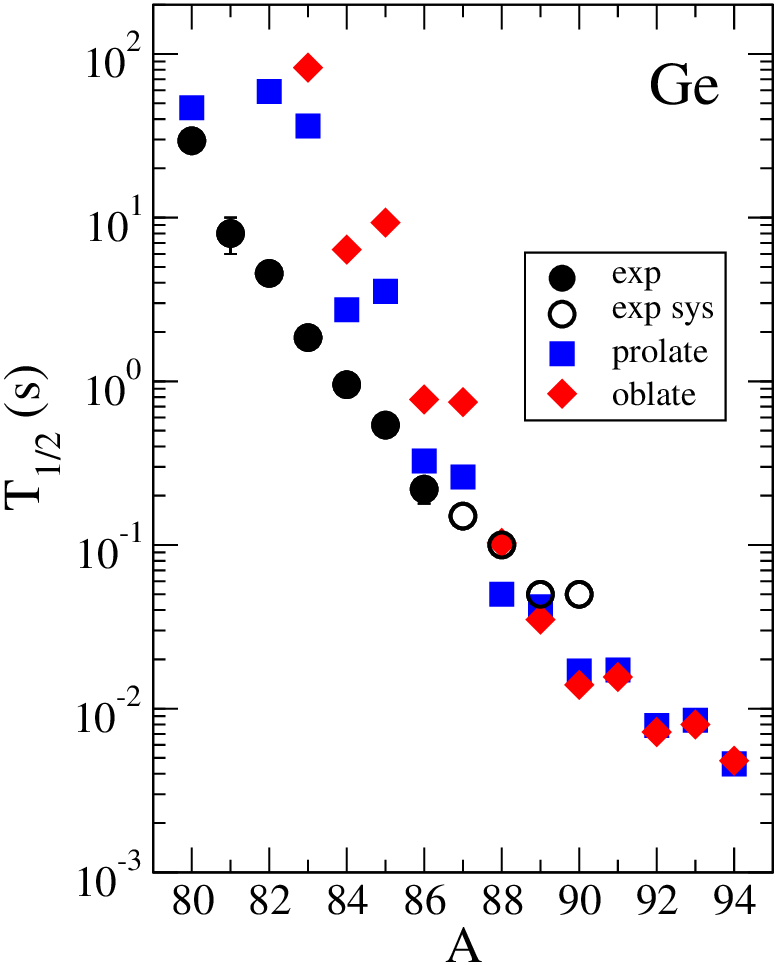}
\caption{(Color online) Measured  $\beta$-decay half-lives for Ge isotopes 
compared to theoretical QRPA-SLy4 results calculated from different shapes. 
Circles are experimental values (open circles are experimental values from 
systematics) \cite{audi12}.}
\label{fig_t12_ge}
\end{figure}

\begin{figure}[ht]
\centering
\includegraphics[width=85mm]{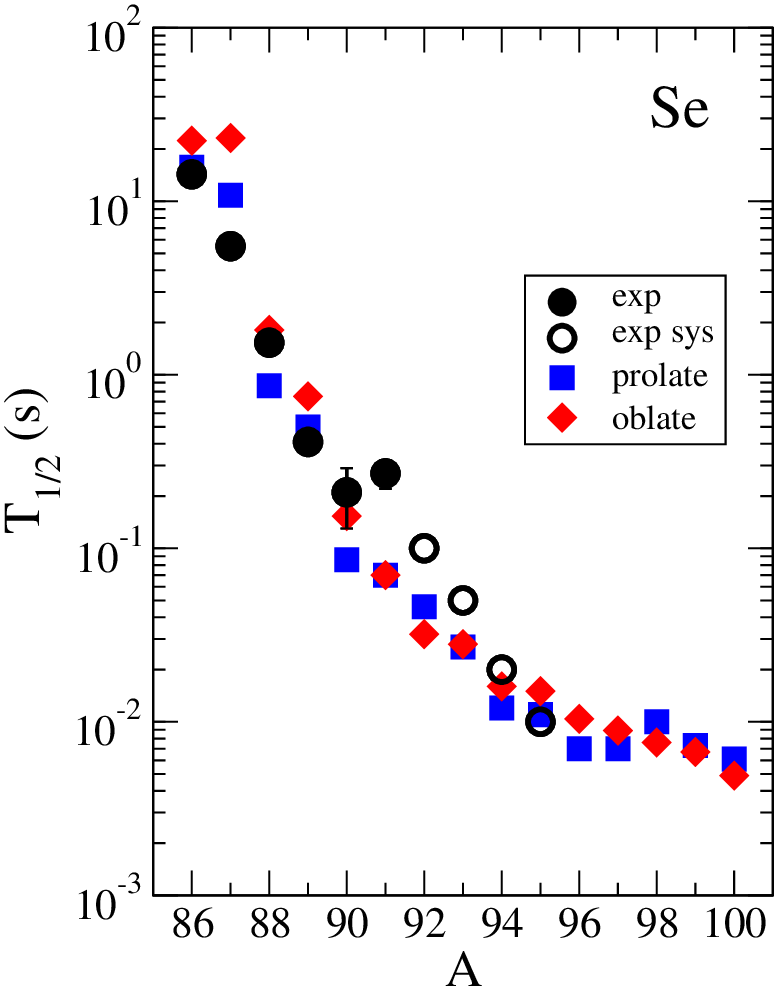}
\caption{(Color online) Same as in Fig. \ref{fig_t12_ge}, but for Se
isotopes.}
\label{fig_t12_se}
\end{figure}

\begin{figure}[ht]
\centering
\includegraphics[width=85mm]{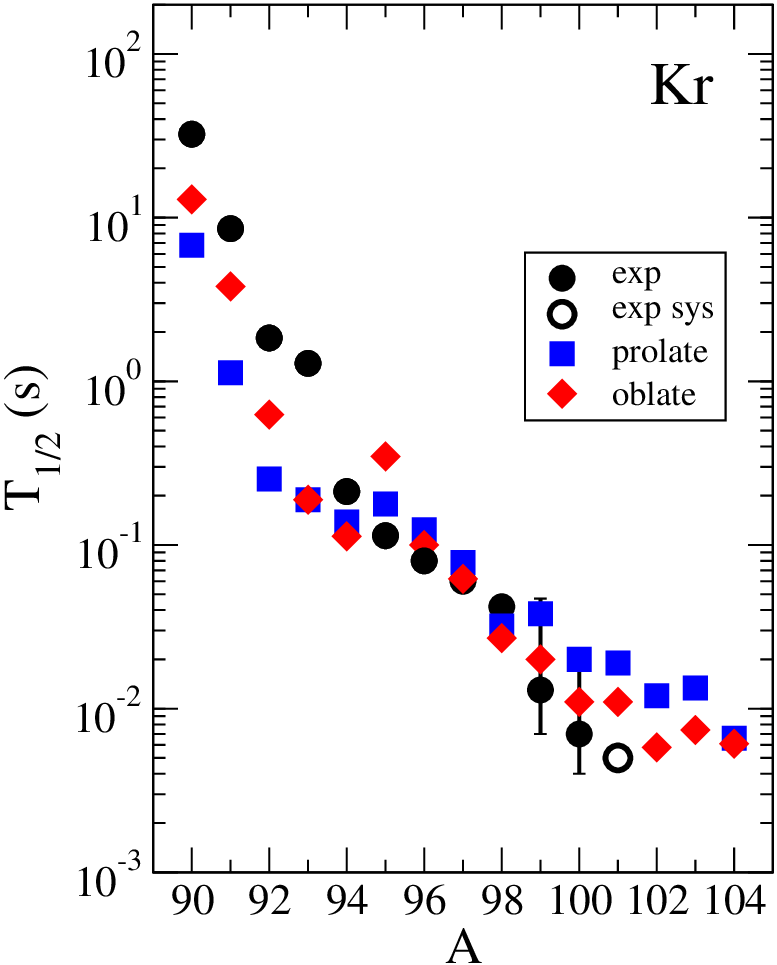}
\caption{(Color online)  Same as in Fig. \ref{fig_t12_ge}, but for Kr
isotopes. Experimental half-lives are from  \cite{audi12,nishimura11}.}
\label{fig_t12_kr}
\end{figure}

\begin{figure}[ht]
\centering
\includegraphics[width=85mm]{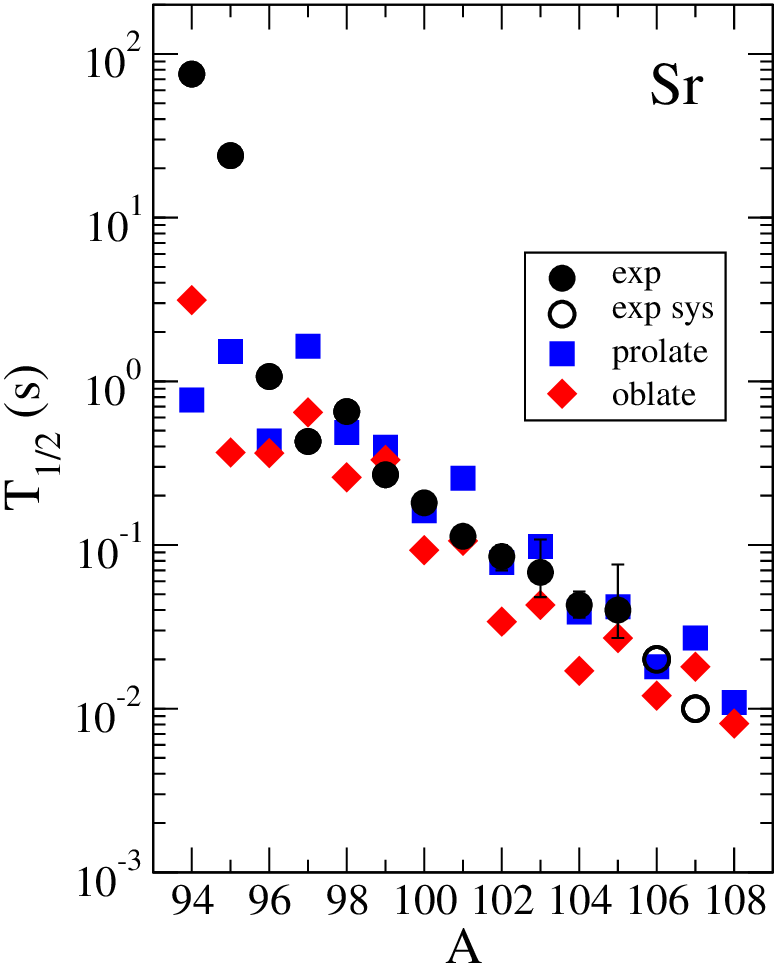}
\caption{(Color online)  Same as in Fig. \ref{fig_t12_ge}, but for Sr
isotopes. Experimental half-lives are from  \cite{audi12,nishimura11}.}
\label{fig_t12_sr}
\end{figure}

\begin{figure}[ht]
\centering
\includegraphics[width=85mm]{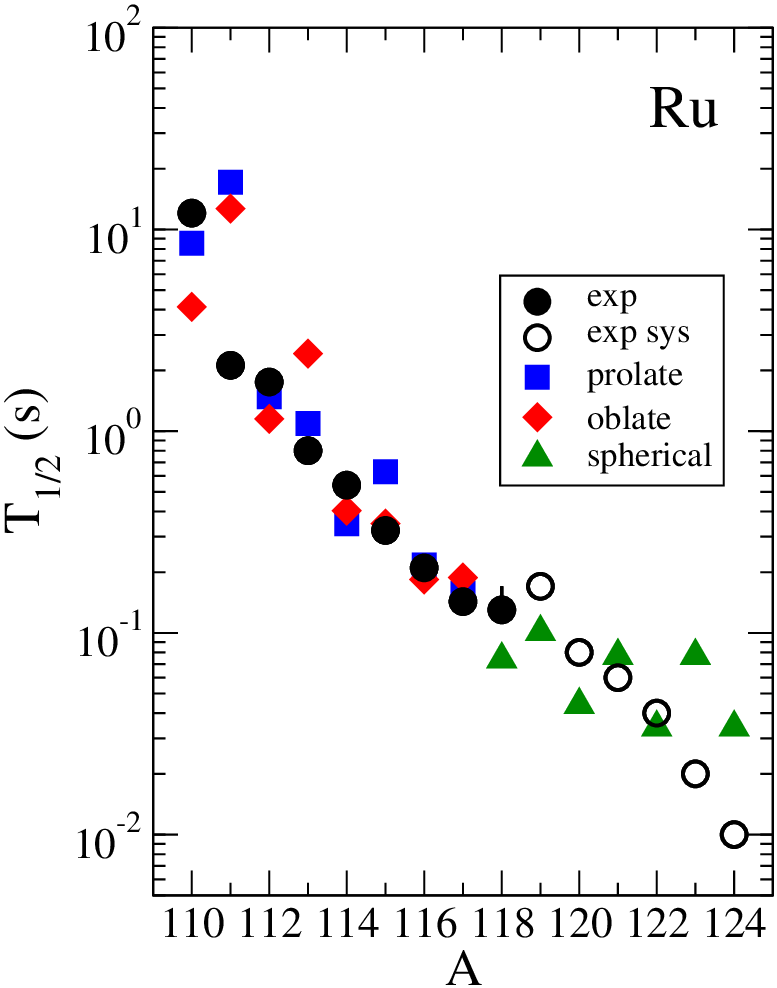}
\caption{(Color online)  Same as in Fig. \ref{fig_t12_ge}, but for Ru
isotopes.}
\label{fig_t12_ru}
\end{figure}

\begin{figure}[ht]
\centering
\includegraphics[width=85mm]{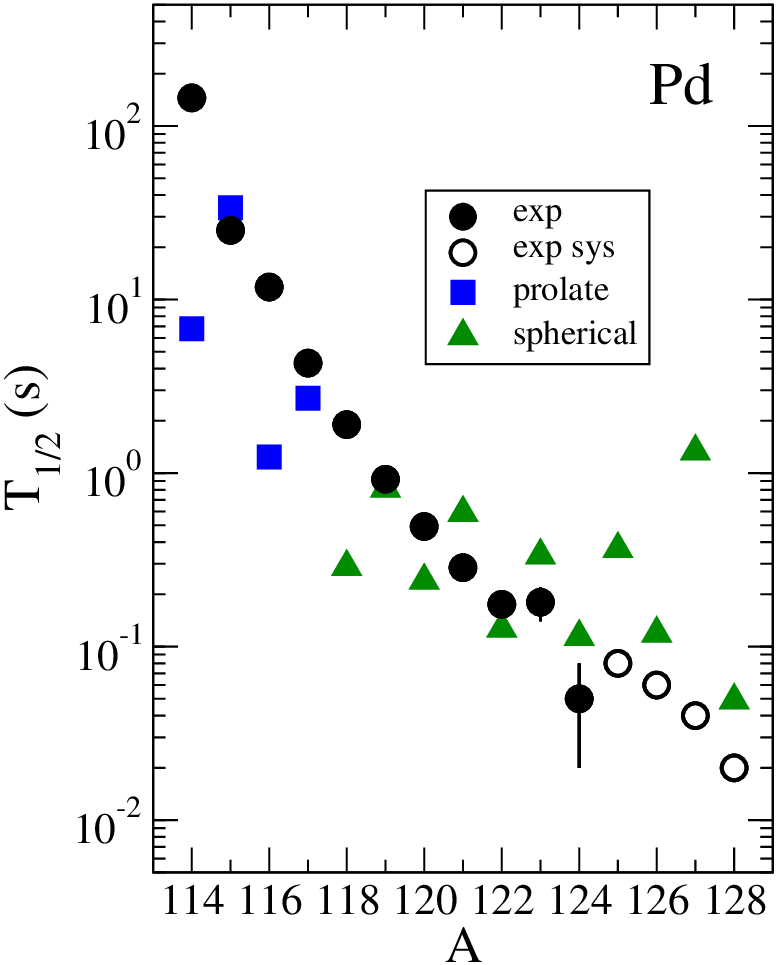}
\caption{(Color online) Same as in Fig. \ref{fig_t12_ge}, but for Pd
isotopes.}
\label{fig_t12_pd}
\end{figure}

\begin{figure}[ht]
\centering
\includegraphics[width=85mm]{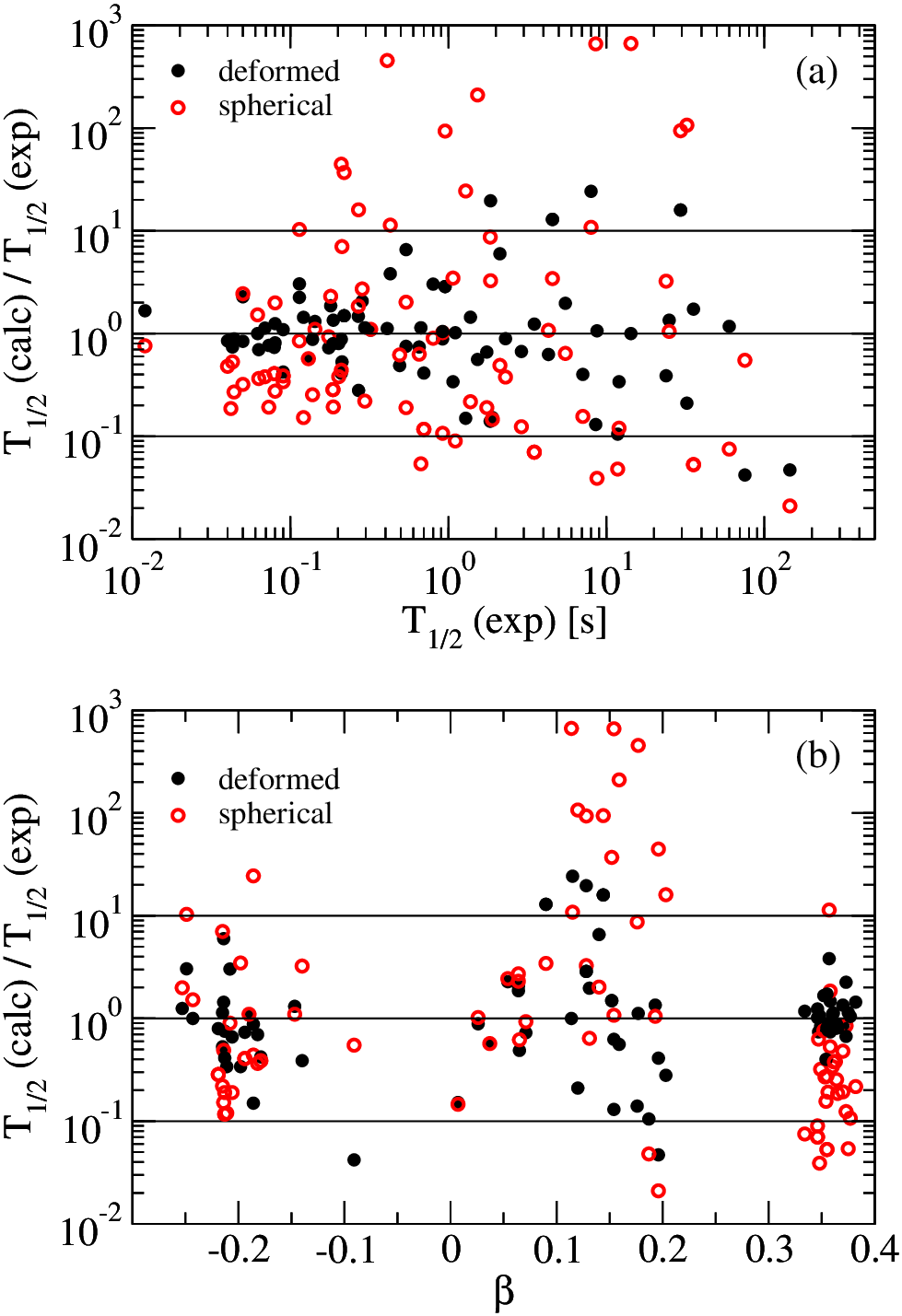}
\caption{(Color online) Ratio of calculated to experimental $\beta$-decay half-lives
for two sets of calculations, with the spherical approximation (open dots) 
and with the deformation that corresponds to the minimum of the PECs (solid dots).
The ratios are plotted as a function of the experimental half-lives (a) and as a 
function of the quadrupole deformation at the minimum of the PECs (b). 
}
\label{fig_ratios}
\end{figure}

\subsection{Gamow-Teller strength distributions}

In the next figures, the energy distributions of the GT strength corresponding 
to the various deformed equilibrium shapes are shown for each isotopic chain.
The results are obtained from QRPA 
with the force SLy4 with pairing correlations and with residual 
interactions with the parameters written in Sec. \ref{sec2}.  The GT 
strength is plotted versus the excitation energy of the daughter 
nucleus with a quenching factor 0.77. Zr and Mo isotopes were already 
studied in Refs. \cite{sarripere,sarri14} and are not repeated here.

Figures \ref{fig_ge_fold}--\ref{fig_pd_fold} contain the results for Ge, 
Se, Kr, Sr, Ru, and Pd isotopes. The energy distributions of the individual 
GT strengths corresponding to the ground state shapes are shown, 
together with continuous distributions for the ground state shapes
as well as for the other possible shapes, obtained 
by folding the strength with 1 MeV width Breit-Wigner functions. 
$Q_\beta$ values are shown with vertical arrows. In both cases,
even and odd isotopes, the $Q_\beta$ values increase with the number of 
neutrons in each isotopic chain and the values in the odd-$A$ isotopes 
$(Z,N+1)$ are about 2-3 MeV larger than the values in the neighbor 
even-even isotopes $(Z,N)$.
The general structure of the GT distributions is characterized by the
existence of a GT resonance, which is placed at increasing excitation 
energy as the number of neutrons $N$ increases in a given isotopic chain.
The total GT strength also increases with $N$, as it is expected to 
fulfill the Ikeda sum rule. 
The various shapes produce quite similar GT strength distributions on 
a global scale. Nevertheless, the small differences among the various 
shapes at the low energy tails (below the $Q_\beta$) of the GT strength 
distributions that can be appreciated because of the logarithmic scale, 
lead to sizable effects in the $\beta$-decay half-lives. 

Unfortunately, comparison with experiment is still not possible for 
the GT strength distributions, the measured half-lives will be compared 
to the calculations in the next subsection. Comparison with calculated 
GT distributions from other theoretical approaches is also restricted 
to the few cases where these results have been published 
\cite{fang13,yoshida13}. In Refs. \cite{fang13,fang10} the authors 
performed QRPA calculations with deformed Woods-Saxon 
potentials and realistic CD-Bonn residual forces using the $G$-matrix 
formalism and compared these results with the results obtained from
separable forces. While in Ref. \cite{fang13} the comparison between 
the results obtained from realistic or separable residual interactions 
is restricted to the half-lives, in Ref. \cite{fang10} the authors
compared those results in the context of two-neutrino double-beta 
decay, concluding that both approaches, realistic and separable, 
lead to similar results.
On the other hand, in Ref. \cite{yoshida13} the Skyrme force SLy4 was
used to generate the mean field as it is done in this work. 
The residual interaction in the ph channel was self-consistently 
introduced and not reduced to a separable form. Finally the pp
residual interaction was written as a contact force with a coupling
strength fitted to reproduce the half-life in $^{100}$Zr.
The GT strength distributions in neutron-rich Zr isotopes obtained
from this approach were compared with the corresponding distributions
obtained with separable forces in Figs. 5-6 in Ref. \cite{yoshida13}. 
From this comparison one can
conclude that in many aspects the main characteristics of the consistent 
force are maintained by a separable force with a much lower computational 
cost. The comparison of the half-lives shows also a remarkable agreement
between both approaches.

In the next figures, Figs. \ref{fig_ge_acum}--\ref{fig_pd_acum}, one
can see in more detail the accumulated GT strength in the energy region 
below the corresponding $Q_\beta$ energy of each isotope, which is the 
relevant energy range for the calculation of the half-lives. 
The vertical solid (dashed) arrows show the $Q_\beta$ ($S_n$) energies, 
taken from experiment \cite{audi12}. In these figures the sensitivity 
of these distributions to deformation can be appreciated and one can 
understand that measurements of the GT strength distribution from 
$\beta$-decay can be, in particular cases, an additional source of 
information about the nuclear deformation, as it was shown in 
Refs. \cite{exppoirier,expnacher,expperez}. 
The GT strength distribution in odd-$A$ isotopes is found to be 
displaced to higher energies (typically about 2-3 MeV) with respect 
to the even-even case.  The shift corresponds roughly to the breaking 
of a neutron pair and therefore it amounts to about twice the neutron 
pairing gap. Below this energy only transitions involving the odd 
nucleon are possible.

The energy distribution of the GT strength is fundamental to constrain 
the underlying nuclear structure. For a theoretical model, it represents 
a more demanding test than just reproducing half-lives or total GT 
strengths that are integral quantities obtained from these strength 
distributions properly weighted with phase factors (see Eq. \ref{t12}). 
These quantities might be reproduced even with wrong strength 
distributions.
This is of especial importance in astrophysical scenarios of high 
densities and temperatures that cannot be reproduced in the laboratory.
Given that the phase factors in the stellar medium are different from 
those in the laboratory, the stellar half-lives become dependent on 
the electron distribution in the stellar plasma that eventually may 
block the $\beta$-particle emission \cite{langanke}.
Therefore, to describe properly the decay rates under extreme conditions 
of density and temperature, it is not sufficient to reproduce the
half-lives in the laboratory. One needs, in addition, to have a reliable 
description of the GT strength distributions \cite{sarri091,sarri092}.

\subsection{Beta-decay Half-lives}

The calculation of the half-lives in Eq. (\ref{t12}) involves knowledge
of the GT strength distribution and of the $\beta$ energies ($Q_\beta-E_{ex}$),
which are evaluated by using $Q_\beta$ values obtained from the mass 
differences between parent and daughter nuclei obtained from SLy4 with a 
zero-range pairing force and Lipkin-Nogami obtained from the code 
HFBTHO \cite{masssly4}.

In Figs. \ref{fig_t12_ge}--\ref{fig_t12_pd} the measured $\beta$-decay 
half-lives (solid dots, open dots stand for experimental values from systematics) 
\cite{audi12,nishimura11} are compared with the theoretical 
results obtained with the prolate, oblate, and spherical equilibrium shapes, 
for the various isotopic chains.
In Fig. \ref{fig_t12_ge} one can see the half-lives for Ge isotopes. The 
lighter isotopes are not well reproduced, being largely overestimated.
This point will be discussed later.
The half-lives obtained from oblate shapes are larger than the corresponding 
prolate ones. This feature is correlated with the GT strength contained 
below the $Q_\beta$ energy in Figs. \ref{fig_ge_fold} and \ref{fig_ge_acum}. 
Prolate shapes, which are closer to experiment, are also the ground states in 
this range of masses according to the calculations (see Fig. \ref{ge_beta_rc}).
For heavier isotopes, the half-lives for oblate and prolate shapes are very 
similar.
In the case of Se isotopes in Fig. \ref{fig_t12_se}, the calculations also 
overestimate the half-lives of the lighter isotopes, but the agreement
with experiment is in this case much better. 
In the middle region the experimental half-lives, 
which are taken from systematics, are reasonably well reproduced. 
The half-lives of heavier isotopes exhibit a rather flat behavior.
Half-lives of Kr isotopes are shown in Fig. \ref{fig_t12_kr}. As in the 
previous figures, the half-lives from the oblate shapes are larger than the 
prolate ones in the lighter Kr isotopes, but the situation is reversed 
at $^{94}$Kr. This is again nicely correlated with the GT strength at low 
excitation energies shown in Fig. \ref{fig_kr_acum}. In general, the 
half-lives in the middle region are well described. This is also true for 
Sr and Ru isotopes 
in Figs. \ref{fig_t12_sr} and \ref{fig_t12_ru}, respectively, where 
the trends observed experimentally are well reproduced, except for the
lighter Sr isotopes that are clearly underestimated and the
heavier Ru isotopes, where the data from systematics fall down faster 
than the calculations. Finally, in the case of Pd isotopes, 
shown in Fig. \ref{fig_t12_pd}, the calculations underestimate 
(overestimate) the measured half-lives in the lighter (heavier) isotopes.

All in all, the agreement with experiment is reasonable, especially 
in the middle regions.
These regions contain in general well deformed nuclei, where the present 
approach is more suitable. On the other hand, weakly deformed 
transitional isotopes, such as light Ge and Se isotopes and heavy 
Ru and Pd isotopes are not so well described. Furthermore, in the light 
isotopes of all the isotopic chains, which are closer to the valley of 
stability, the half-lives are larger because of the small $Q_\beta$ 
energies involved. In these cases the half-lives are determined 
exclusively by the very low energy tail of the GT strength distribution 
contained in the narrow window below $Q_\beta$. Therefore, tiny variations 
in the description of the GT strength distribution in the low-lying energy 
region can drive sizable effects in the half-lives. Of course it is also 
important to describe the half-lives of the long-lived isotopes, but 
their significance to constrain the GT strength distribution is minor
since the half-lives are insensitive to most of this distribution.

Half-lives for neutron-rich Kr, Sr, Zr, and Mo isotopes calculated from 
self-consistent deformed QRPA calculations with the Gogny D1M interaction
and experimental values of $Q_\beta$ \cite{peru14} agree with the results 
in this work within the uncertainties of the calculations.
The agreement is also very reasonable between the calculated half-lives and 
those obtained from deformed QRPA calculations using deformed Woods-Saxon 
potentials to generate the mean field and complemented with realistic 
CD-Bonn residual forces \cite{fang13,ni14}. 
The agreement is also good with the results in Ref. \cite{yoshida13}
using the Skyrme force SLy4 with consistent residual interactions in 
the ph channel as mentioned earlier. Fig. 7 in that reference displays
this comparison.

It is also worth noticing that the worst agreement with experiment
occurs in the light Ge isotopes, as well as in heavy Pd isotopes. In these 
cases the calculations overestimate the experiment leaving 
room for contributions coming from first forbidden (FF) transitions.
One can understand from simple qualitative arguments that the 
role of FF transitions is expected to be more important in 
lighter Ge and in Pd isotopes.
Thus, for $_{32}$Ge, $_{34}$Se, $_{36}$Kr, and $_{38}$Sr isotopes, the 
last occupied proton orbitals come basically from the $2p_{3/2},1f_{5/2}$ 
and $2p_{1/2}$ negative-parity spherical shells.
On the other hand the neutrons in $^{80-94}$Ge isotopes occupy
orbitals belonging to the $1g_{9/2},2d_{5/2}$ and $1g_{7/2}$ 
positive-parity spherical shells.
Therefore, in the $\beta$-decay one neutron in a positive-parity
state is transformed into a proton that would sit in a 
negative-parity state, thus suppressing GT and favoring FF transitions
in the low-lying transitions. 
This is particularly true for the lighter Ge isotopes. In the heavier
ones, other neutron states with negative parity ($1h_{11/2}$) have
to be considered because of deformation effects.
The same argument can be applied to the lighter Se, Kr, and Sr 
isotopes, but in these cases proton states from positive parity 
($1g_{9/2}$) are closer in energy and would participate in the decay 
favoring GT transitions.
The situation is different in the case of Ru and Pd isotopes. Now 
the available proton states for the decay are of positive parity 
($1g_{9/2}$), whereas most of the last occupied neutrons belong to 
negative-parity states ($1h_{11/2}$), thus favoring FF transitions.
According to calculations \cite{borzov3,fang13} of the FF transitions 
in this mass region, minor effects are expected from them. Nevertheless, 
it would be very interesting in the future to study systematically the 
FF contributions in all the isotopes in this mass region.

Another feature observed in the present calculations is the existence of some
odd-even staggering effect in the calculated half-lives, which is not 
observed experimentally. This effect is particularly evident in 
Ru and Pd isotopes. There are not many calculations involving 
simultaneously even-even and odd-$A$ isotopes, but some of them 
exhibit some sort of staggering effect as well \cite{fang13}.
The appearance of this effect in the half-lives suggests
some deficiency in the model that might be related to the 
determination of ground-state energies in the odd-$A$ systems
\cite{duguet01}. 
Unfortunately, there are more sources of uncertainty related to the
odd-$A$ systems that should be considered as well \cite{schunk10}, such 
as the spin and parity assignments, the blocking procedures or the
treatment of the 1qp excitations involving the odd nucleon. 
This issue will be the subject of a future investigation in this 
direction.

It is also interesting to look for the simultaneous appearance of structural 
effects that eventually can appear in different observables. One example
can be seen in the evolution of the experimental
half-lives with the number of neutrons in the isotopic chains. 
At some points one observes discontinuities in the general 
trends of behavior, such as in the mass regions $^{90,92}$Se, 
$^{92,94}$Kr, $^{96,98}$Sr, and $^{118,120}$Ru. These experimental
findings on the half-lives are correlated with the shape transitions 
in Figs. \ref{se_beta_rc}--\ref{ru_beta_rc} predicted in the model.
One cannot state firmly that these sharp changes in the behavior of
the half-lives are signatures of shape transitions, but
certainly this correlation cannot be discarded given that
a change of the deformation in the nuclear system involves
a structural change to whom the half-lives are also sensitive.

Finally, the impact of deformation on the decay properties can be 
better appreciated in a systematic comparison of the half-lives
calculated with both the spherical approximation and the deformation
that corresponds to the minimum of the PEC for each isotope.
Then,  Fig. \ref{fig_ratios} shows the ratios of the calculated
and experimental half-lives for two sets of data corresponding to a
spherical calculation (open dots) and to a deformed calculation 
(solid dots) at the self-consistent deformation that gives the minimum 
of the PECs. These ratios are plotted as a function of the experimental 
half-lives (a) and as a function of the quadrupole deformation at the 
minimum of the PECs (b). 
To increase the size of the sample, besides the isotopes considered in 
this work with measured half-lives, I have also included the set of Zr 
and Mo neutron-rich isotopes studied in Ref. \cite{sarri14} with
measured half-lives.
In the upper panel of Fig. \ref{fig_ratios} (a) one can see how deformation
improves the description of the half-lives. Practically all the full black 
dots are contained within the horizontal lines defining the region of
one order of magnitude agreement. On the other hand, the results from
the spherical calculation are more spread out with larger discrepancy
with experiment. One can also see that the results are better in both
spherical and deformed calculations for shorter half-lives, 
whereas the results for larger half-lives show sizable deviations.
The latter correspond to isotopes close to the valley of stability
with small $Q_\beta$-values, where the half-lives are only sensitive
to the small portion of the GT strength distribution at low excitation
energies below $Q_\beta$.
In the lower panel (b) one can see the results from a different point of view
and it can be studied whether deformation improves the results evenly in the
whole range of deformations or whether its effect is stronger at large
deformations. Three regions of accumulation of results can be distinguished. 
Two of them correspond to well deformed nuclei located
at $\beta \approx -0.2$ and $\beta \approx 0.35$. In these regions the
deformed calculations clearly improve the results from the spherical ones
that show a tendency to underestimate the experiment. The other region
corresponds to $0 < \beta < 0.2$ values, where nuclei are softer or
transitional and the deformed formalism should be improved. In this
region the results are more scattered than in the well deformed regions,
but the deformed calculations show deviations that rarely exceed one order
of magnitude, still representing an improvement over the spherical results.

In order to have a quantitative estimation of the quality of the various
calculations, following the analysis made in Ref. \cite{moller3}, the 
logarithms of the ratios of the calculated and experimental half-lives 
are introduced through the quantities 

\begin{equation}
r=\log_{10} \left[ \frac{T_{1/2} ({\rm calc})}  {T_{1/2} ({\rm exp})} \right] \, .
\end{equation}
Then, the average position of the points, $M_r$, the standard deviation, 
$\sigma_r$, and the total error, $\Sigma_r$, are defined as

\begin{eqnarray}
M_r= \frac{1}{n}\sum_{i=1}^n  r_i\, ; 
\quad
\sigma_r&=&\left[ \frac{1}{n}\sum_{i=1}^n \left( r_i -M_r \right) ^2 \right]^{1/2}\, ;
\nonumber  \\
\Sigma_r&=&\left[ \frac{1}{n}\sum_{i=1}^n \left( r_i \right) ^2 \right]^{1/2}\, ,
\end{eqnarray}
and their corresponding factors   
$M_r^{10}=10^{M_r}$, $\sigma_r^{10}=10^{\sigma_r}$, and $\Sigma_r^{10}=10^{\Sigma_r}$.
The analysis of the
results shown in Fig. \ref{fig_ratios} involving $n=81$ nuclei leads to the
values $M_r^{10}=1.105$, $\sigma_r^{10}=10.21$, and 
$\Sigma_r^{10}=10.24$ in the spherical case and  
$M_r^{10}=0.937$, $\sigma_r^{10}=3.09$, and
$\Sigma_r^{10}=3.09$ in the deformed one, showing clearly the improvement 
achieved with the deformed formalism.

\section{CONCLUSIONS}

A microscopic approach based on a deformed QRPA calculation on 
top of a self-consistent mean field obtained with the SLy4 Skyrme 
interaction has been used to study the nuclear structure and 
the decay properties of even and odd neutron-rich isotopes in 
the mass region $A\approx 80-130$. The nuclear model and interaction 
have been successfully tested in the past providing good agreement 
with the available experimental information on bulk properties all 
along the nuclear chart. Decay properties in different mass regions 
have been well reproduced as well. The structural isotopic evolution 
has been studied from their PECs. Depending on the isotopic chain, 
a large variety of nuclear shapes is found, including spherical shapes,
well developed deformed shapes, and transitional soft shapes.
Charge radii have been also investigated, showing the connection
between a discontinuous behavior in the isotopic trend with a
shape transition and comparing the results with the available
measurements from laser spectroscopy. Then, 
Gamow-Teller strength distributions and $\beta$-decay half-lives have 
been computed for the equilibrium shapes.

The isotopic evolution of the GT strength distributions exhibits some
typical features, such as GT resonances increasing in energy and 
strength as the number of neutrons increases. Effects of deformation
are hard to see on a global scale, but they become apparent in the 
low excitation energy below $Q_\beta$ energies, a region that determines
the half-lives. Half-lives have been calculated using $Q_\beta$ 
energies calculated with the force SLy4. 
In general, a reasonable agreement with experiment is obtained, especially in 
the short-lived nuclei of Ge, Se, Kr, Sr, and Ru isotopes. The results
are comparable to other calculations using different approaches for the
mean field and/or residual interactions. Special difficulties are found to 
describe properly the half-lives of the lighter Ge isotopes and the Pd 
isotopes. These are examples of transitional 
nuclei where the nuclear structure is more involved and the concept of 
a well defined shape might not be meaningful. 

A systematic comparison of the ratios of the calculated and experimental
half-lives has been done using both spherical and deformed calculations,
showing that the inclusion of deformation improves significantly the
description of the decay properties.

Experimental information on the energy distribution of the GT strength 
is a valuable piece of knowledge about nuclear structure in this mass 
region. The study of these distributions is within the current 
experimental capabilities in the case of the lighter isotopes considered 
in this work. Here, I have presented theoretical predictions for them
based on microscopic calculations. 
Similarly, measuring the half-lives of the heavier isotopes will be highly 
beneficial to model the {\it r} process and to constrain theoretical 
nuclear models. This possibility is also open within present capabilities
at RIKEN.

\begin{acknowledgments}
This work was supported by Ministerio de Econom\'\i a y Competitividad
(Spain) under Contract No. FIS2011--23565 and the 
Consolider-Ingenio 2010 Program CPAN CSD2007-00042.
\end{acknowledgments}

\end{document}